# Magnetic molecular wheels and grids - the need for novel concepts in "zero-dimensional" magnetism

Oliver Waldmann*

Department of Chemistry and Biochemistry, University of Bern, Freiestrasse 3, CH-3012 Bern, Switzerland.

**Contents**

* *E-mail address:* waldmann@iac.unibe.ch (O. Waldmann)

**Abstract**

Supramolecular chemistry has allowed the production, by self-assembly, of inorganic complexes with a [N × N] square matrix-like configuration of $N^2$ metal centers. Interest in these systems is driven by the potential applications in information technology suggested by such a "two-dimensional" (2D), addressable arrangement of metal ions. From the magnetic perspective, [N × N] grids constitute molecular model systems for magnets with extended interactions on a square lattice, which have gained enormous attention in the context of high-temperature superconductors. Numerous [2 × 2] grids as well as a few [3 × 3] grids with magnetic metal ions such as Cu(II), Ni(II), Co(II), Fe(II), and Mn(II) have been created. Magnetic studies unraveled a remarkable variety in their magnetic properties, which will be reviewed in this work with emphasis on the underlying physical concepts. An intriguing issue is the connection of [2 × 2] and [3× 3] grids with "one-dimensional" (1D) rings, as experimentally realized in the molecular wheels. For a [2 × 2] square of spin centers the distinction between a 2D grid and a 1D ring is semantic, but also a [3 × 3] grid retains 1D character: it is best viewed as an octanuclear ring with an additional ion "doped" into its center. Challenging familiar concepts from conventional magnets, the current picture of elementary excitations in antiferromagnetic rings will be discussed, as a prerequisite to understand the complex [3 × 3] grids.

**1. Introduction**

In the last decade the investigation of nano-sized magnetic molecules such as the celebrated Mn12 molecule has evolved into one of the most attractive research areas in the field of molecule-based magnets. Several excellent reviews are available documenting the many merits and the chemical, physical, and technological implications of these molecular nanomagnets (MNMs) [1-9]. This work thus does not elaborate on such issues, but instead explores MNMs from a physical point of view, asking what magnetic phenomena may arise and which physical concepts allow one to rationalize them.

The magnetic properties of MNMs are in principle simple to describe. The basic principles were established decades ago and nothing essential needs to be added here. For instance, the discussion for most MNMs of current interest starts with the spin Hamiltonian

$$\hat{H} = -\sum_{i \neq j} J_{ij} \hat{\mathbf{S}}_i \cdot \hat{\mathbf{S}}_j + \sum_i \hat{\mathbf{S}}_i \cdot \mathbf{D}_i \cdot \hat{\mathbf{S}}_i + \mu_B \sum_i g_i \hat{\mathbf{S}}_i \cdot \mathbf{B} , \qquad (1)$$

which consists of the isotropic interaction terms due to Heisenberg exchange, the single-ion anisotropy terms due to ligand-field interactions, and the Zeeman term [10]. $\hat{\mathbf{S}}_i$ is the spin operator of the i-th ion with spin $S_i$, $\mu_B$ is the Bohr magneton, and **B** the magnetic field. The approximations in Hamiltonian (1), such as assuming magnetic ions with orbitally non-degenerate ground states or neglecting dipole-dipole and anisotropic interactions [10], are obvious and shall not to be discussed further here. In the following a Hamiltonian such as Hamiltonian (1) will be referred to as a microscopic spin Hamiltonian in order to clearly distinguish it from effective models, which appear later on.

Hamiltonian (1) is in fact sufficient to explore the main characteristics of MNMs. On a fundamental level, the magnetic phenomena encountered in MNMs arise from the mutual competition of the three elementary magnetic contributions: magnetic interaction, magnetic anisotropy, and applied magnetic fields. Hamiltonian (1) constitutes a generic model in this regard, as all three effects are represented by their simplest possible terms. Situations of course arise where the spin terms neglected in Hamiltonian (1), and couplings to environmental degrees of freedom, cause important effects. For example, in the single-molecule magnet (SMM) Mn12, the tunneling splitting is small, on the order of $10^{-7}$ K, and the tunneling dynamic is thus strongly affected by e.g. nuclear hyperfine fields. The defining characteristic of the Mn12 molecule, however, is a large S = 10 ground-state spin with a pronounced easy-axis anisotropy. This allows one to treat it just as a single, large spin (at least as long as only low-temperature properties are concerned) and to disregard the underlying many-spin nature of the system entirely. In fact, most experimental findings can be convincingly discussed within this "giant spin" model, or, in physical terms, by an effective spin Hamiltonian which resembles the one used for ordinary paramagnets, for instance by

$$\hat{H}_S = D\hat{S}_z^2 + B_{40}\hat{O}_4^0 + B_{44}\hat{O}_4^4 + \mu_B g \hat{\mathbf{S}} \cdot \mathbf{B} . \qquad (2)$$

Despite its success, such an approach obviously does not unravel, for example, why Mn12 exhibits a S = 10 ground state. Any attempt to understand Mn12, or MNMs in general, from a fundamental point of view thus starts with Hamiltonian (1).

At this point some general question are perfectly obvious: i) What kind of magnetic phenomena may arise from Hamiltonian (1), ii) under what conditions do they arise, and, equally importantly, iii) how can they be rationalized in physical terms. The variables in this "game" are the coupling matrix $J_{ij}$ and the vectors $\mathbf{D}_i$, $g_i$, and $S_i$ (matrices and vectors with respect to the site index i, j) and the problem is to find, and understand, the possible outcomes. Playing this game is not only of physical interest. For instance, to know the manifold of microscopic parameters for which Hamiltonian (1) produces a large ground-state spin with easy-axis anisotropy should help synthetic chemists to devise strategies to eventually create SMMs with higher blocking temperatures.

In general, the above task is difficult, and different strategies have been used to tackle it. For most MNMs studied so far, the Heisenberg exchange term is much stronger than the magnetic anisotropy. It is thus natural to consider first the situation with isotropic interactions alone, and then the effects of a weaker magnetic anisotropy. In the traditional approach the anisotropy is treated in first-order perturbation theory (the so called strong exchange limit). However, it has become increasingly clear in recent years that many interesting effects are not grasped by this approach, and more sophisticated techniques are required. One such technique, which is relied on heavily in this work, is to combine numerical calculations for Hamiltonian (1) and the effective spin Hamiltonian approach, in order to uncover first general physical trends and then to devise effective spin Hamiltonians, which cover the essential physics of the system (the numerical calculation of magnetic properties by itself is not of concern here).

This work focuses on [2 × 2] and [3 × 3] grid molecules and antiferromagnetic molecular wheels. Grid and ring molecules have a great deal in common. Both systems exhibit very high molecular symmetry. Accordingly, the number of free parameters in the microscopic Hamiltonian is considerably reduced. Even if the assumed high symmetry is only approximately realized in the actual molecule, the variation in the parameters due to the deviation from perfect symmetry is small and difficult to resolve in experiment. Thus these systems are generally described to a very high degree of accuracy by the simplest model Hamiltonian consistent with their structure and the maximal symmetry. For instance, their symmetrical, planar structure dictates uniaxial magnetic anisotropies (with the anisotropy axis, denoted as the z axis, perpendicular to the plane of the molecules). In other words, they are excellent model systems to play the above game. Grid and ring molecules are also related from another perspective. The distinction between a [2 × 2] grid and a tetra-nuclear ring, or a square, is obviously semantic. This actually leads to some ambiguity in the classification of tetra-nuclear molecules as [2 × 2] grids, rings, or squares - a point which shall not be considered here. For [3 × 3] grids the situation is subtler. For instance, the so called Mn-[3 × 3] grid was shown to be best described as an octanuclear ring with an additional ion "doped" into its center. With this analogy and the results on the molecular wheels at hand, the magnetism of this difficult system can be successfully elucidated.

The aim of this work is twofold. On the one hand, it aims at reviewing the magnetic phenomena provided by the grid molecules, demonstrating the broad range of possible effects realized in such, conceptually simple, model systems. The isotropic coupling situation encountered in grids has been reviewed extensively [11] and is not of much concern here. Instead, those effects will be emphasized which are beyond a pure isotropic exchange model, involving e.g. magnetic anisotropy as embodied in Hamiltonian (1). On the other hand, recent advances in the understanding of the spin structures, or, in a physical language, of the nature of the elementary excitations in antiferromagnetic spin clusters will be expounded. This unfolds physical concepts of general relevance; with the Mn-[3 × 3] grid as a particular example.

In the next section the magnetism of the [2 × 2] grid molecules is explored. Subsequently, the nature and physical description of the elementary excitations in antiferromagnetic Heisenberg rings are outlined, pinpointing physical ideas of general interest. In the fourth section, the magnetism of the [3 × 3] grid molecules, with emphasis on the Mn-[3 × 3] grid, is discussed. In the last section, the generality of the concepts developed for the rings and related systems, and conclusions thereof, will be examined. Two extensive reviews on [N × N] grid molecules are available, covering the ligand systems, the resulting grid complexes, chemical aspects, and the electrochemical, optical, and some magnetic properties, as well as conclusions concerning applications [11,12].

## 2. Magnetic phenomena in [2 × 2] grids

Many complexes with a planar, square matrix like arrangement of four paramagnetic metal centers, such as Cu(II), Ni(II), Co(II), Fe(II), or Mn(II), have been synthesized so far [11-13]. As indicated in the introduction, classification as [2 × 2] grids, tetranuclear rings, or squares is not always obvious, but as a common feature one may demand a (approximate) molecular $S_4$ or $C_4$ symmetry axis. For simplicity these complexes will be referred to indifferently as [2 × 2] grids henceforth. Magnetic studies have shown a remarkable variety in their magnetic properties, which will be reviewed in this paragraph.

In view of their symmetry, the magnetism of [2 × 2] grids is excellently described by the spin Hamiltonian

$$\hat{H} = -J\left(\sum_{i=1}^{3} \hat{\mathbf{S}}_i \cdot \hat{\mathbf{S}}_{i+1} + \hat{\mathbf{S}}_4 \cdot \hat{\mathbf{S}}_1\right) + D\sum_{i=1}^{4} \hat{S}_{i,z}^2 + \mu_B g_{xy}\left(\hat{S}_x B_x + \hat{S}_y B_y\right) + \mu_B g_z \hat{S}_z B_z , \quad (3)$$

which is obtained as an adaptation of Hamiltonian (1) to the situation of the [2 × 2] grids, with the total spin operator $\hat{\mathbf{S}} = \sum_i \hat{\mathbf{S}}_i$ (here and in the following, constant terms such as $D\sum_i [-S_i(S_i+1)/3]$ are suppressed). A diagonal exchange term $-J_2\left(\hat{\mathbf{S}}_1 \cdot \hat{\mathbf{S}}_3 + \hat{\mathbf{S}}_2 \cdot \hat{\mathbf{S}}_4\right)$ can be neglected as the [2 × 2] grids lack corresponding ligand bridges. Accordingly, four magnetic parameters are sufficient to describe the magnetic properties: the coupling constant J, the zero-field splitting parameter D, and the two g factors $g_{xy}$ and $g_z$. The parameterization $g = \sqrt{(2g_{xy}^2 + g_z^2)/3}$ and $\Delta g = g_z - g_{xy}$ will be also used.

The exchange coupling is generally observed to be antiferromagnetic, $J < 0$ [11,13]; ferromagnetic exchange, $J > 0$, was found in Cu(II)-[2 × 2] grids, and a Ni(II) grid. An example of a Cu(II)-[2 × 2] grid with intra-molecular ferromagnetic exchange is the complex $[Cu_4(6POAP\text{-}H)_4](ClO_4)_4$ (**1**), Fig. 1, which exhibits approximate $S_4$ symmetry [14]. The variation of the effective moment $\mu = \sqrt{3k_B\chi T / N_A\mu_B^2}$ as a function of temperature shows an increase in moment with decreasing temperature from 3.6 $\mu_B$ at room temperature to a maximum of 4.7 $\mu_B$ at 5 K (Fig. 2), which is typical for a square of spin-1/2 centers with ferromagnetic exchange ($k_B$ is the Boltzman and $N_A$ the Avogadro constant). The data could be reproduced with Eq. (3) yielding g = 2.060(7), and J = 27(1) K ($S_i$ = 1/2, D = $\Delta$g = 0). In the fit, the presence of a small amount of impurity, weak intermolecular interactions, and temperature independent paramagnetism was additionally allowed for. The ferromagnetic exchange in this compound, and several others like it [15,16], could be rationalized by their particular structure. The square-pyramidal coordination spheres of the Cu(II) centers lead to $d_{x2-y2}$ ground states with the orbital lobes directed along the short bonds, defining equatorial planes which lie perpendicular to the grid consistent with the (approximate) $S_4$ symmetry of the cluster. The exchange through the alkoxide bridges occurs along short equatorial and long axial contacts, resulting in orthogonal magnetic orbitals and hence the ferromagnetic behavior.

Another interesting situation is encountered in the Ni(II) grid complex $[Ni^4L^b{}_8]\cdot 4CH_2Cl_2$ (**2**) with $L^b = C_5H_4N\text{-}CON\text{-}CN_4\text{-}C_2H_5$, Fig. 3 [17]. This compound crystallizes in the tetragonal space group I4(1)/a, and the cluster thus exhibits crystallographically imposed $S_4$ symmetry. Magnetic studies on powder samples furthermore revealed a sizeable ferromagnetic coupling of the four Ni(II) metal centers within a molecule. Most interestingly, a second species, $[Ni^4L^c{}_8]\cdot 4CH_2Cl_2$ (**3**) with $L^c = C_5H_4N\text{-}CSN\text{-}CN_4\text{-}C_2H_5$, could also be synthesized, in which the ligand differs by only one atom; the oxygen in $L^b$ is replaced by a sulfur. These two $Ni_4$ squares are not only isostructural, but their magnetically relevant geometrical dimensions differ by less than 3.5%. This leads to a novel situation: differences in the magnetic properties originate predominantly from the differences in the electronic properties of oxygen and sulfur, and not from structural differences as investigated in magneto-structural correlations. One may thus speak here of magneto-electronic correlations. The magnetism of both complexes **2** and **3** were investigated in detail by magnetization and high-field torque magnetometry on single crystals [18]. Employing a novel data analysis scheme, this allowed a precise determination of all the parameters in Eq. (3), yielding J = 0.9(1) K, D = -2.7(1) K for **2**, and J = 0.9(1) K, D = -2.0(1) K for **3** [in both clusters, $S_i = 1$, g = 2.2(1), Δg = 0.01(1)]. As the geometry of the Ni coordination spheres are essentially identical, variations in D should be ascribed to different electronic environments of the Ni centers. The observed variation can be rationalized by the different donor capabilities of oxygen and sulfur. Surprisingly, the coupling constants are identical within experimental error for both complexes, although the oxygen/sulfur atoms are expected to participate in the exchange. Obviously, the ferromagnetic coupling in **2** and **3** is the result of a subtle balance of various exchange contributions, which is not easily reconciled.

A broad class of [2 × 2] grids exhibiting exceptional magnetic properties is based on the ligand system L = bis(bipyridyl)pyrimidine (Fig. 4). It coordinates with various divalent metal ions into $M^{II}_4L_4$ grid structures with approximate $D_{2d}$ symmetry, and may be derivatized by various end-groups at the three positions $R_1$, $R_2$, and $R_3$. This enables a wide variety of physical properties, which can be tuned by the choice of the substituents at these positions. From the magnetic perspective, the end-group $R_1$ is expected to be of particular relevance as it modifies the bridging pyrimidine group, which mediates the exchange interaction between neighboring metal centers. As a further interesting aspect of this class of [2 × 2] grids, the mononuclear analogues $[M^{II}(\text{terpyridine})_2]^{2+}$ are available, which exhibit metal coordination spheres very close to that in the [2 × 2] grids. A study on these complexes allows one to gain independent insight on the magnetic anisotropy, whereby providing an unusual view on the exchange situation in the corresponding $M^{II}_4L_4$ grid structures [19].

The magnetism of the grid $[Ni_4(\mathbf{4a})_4](PF_6)_8$ (**5**) was studied in detail by means of magnetization measurements on both powder and single-crystal samples [20]. The temperature dependent data recorded at several magnetic field strengths with orientations perpendicular and parallel to the plane of the molecule could be well reproduced with Hamitonian (3) ($S_i = 1$), but small, significant deviations remained. Further spin terms such as next-nearest-neighbor exchange, anisotropic exchange, antisymmetric exchange, and biquadratic exchange were accordingly considered in the analysis in addition to the terms present in Hamiltonian (3). Unambiguous evidence for a biqudratic exchange, which adds a term $J'\sum_{i=1}^{4}\left(\hat{\mathbf{S}}_i \cdot \hat{\mathbf{S}}_{i+1}\right)^2$ to Hamiltonian (3), was obtained in this way with best-fit parameters J = -8.4 K, J' = 0.51 K, D = -7.9 K, $g_z = 2.10$, and $g_{xy} = 2.02$ (Fig. 5). The complex **5** is a rare example of a system, where biquadratic exchange could be convincingly demonstrated by (careful) magnetization studies.

Initial studies of the powder magnetic susceptibility as function of temperature for the Co(II)-[2 × 2] grid $[Co_4(\mathbf{4a})_4](PF_6)_8$ (**6**) revealed a sizeable intra-molecular antiferromagnetic exchange interaction, and results on the mononuclear analogue $[Co(terpyridine)_2](PF_6)_2$ (**7**) suggested the presence of a pronounced magnetic anisotropy at low temperatures [19]. Subsequently recorded magnetization curves on single crystals of **6** at low temperatures unraveled a remarkable magnetic behavior [21]. The magnetization curves at 1.9 K, recorded at three mutual perpendicular orientations of the magnetic field, exhibit two characteristic features (Fig. 6a): (i) For magnetic fields perpendicular to the grid plane, the magnetization curve shows a behavior reminiscent of a thermally broadened magnetization step due to a ground-state level crossing, as it is often observed in antiferromagnetic clusters [22]. (ii) The magnetic moments for fields in the plane of the grid, in contrast, increase linearly with equal slope as function of magnetic field. The slight deviations visible at higher fields were explained by small misalignments of the crystal with respect to the magnetic field.

The magnetism of high-spin Co(II) centers ($S_i = 3/2$), as in the grid complex **6**, is notoriously difficult to describe because of orbital contributions [23]. However, a careful analysis of the magnetism of the mononuclear compound **7** as well as general theoretical considerations showed that Hamiltonian (3) provides an appropriate description of the situation in **6**. Hamiltonian (3) indeed reproduces the above two characteristics for $|J| << |D|$ and $J, D < 0$ (Fig. 6b). This range of parameter values was also inferred from the experimental findings on both the grid and the terpyridine complex, yielding $J \approx -1.5$ K and D values on the order of several tens of Kelvins.

The magnetization curves observed in the Co(II) )-[2 × 2] grid **6** were explained by a perturbational analysis of Hamiltonian (3). At low temperatures, $k_BT << |D|$, only the lowest-lying Co(II) Kramers duplets are populated, and effective spins $S'_i = 1/2$ were introduced to describe them. The result of first-order perturbation theory is most easily obtained by the substitution $\hat{S}_{i,\alpha} = (g'_\alpha / g_\alpha)^2 \hat{S}'_{i,\alpha}$ for each $\alpha$ = x, y, z [23]. Since $D < 0$, the low-lying Kramers duplets consist of the states with magnetic quantum numbers $m_i = \pm 3/2$, and the effective g factors become $g'_{xy} = 0$ and $g'_z = 3g_z$. Including second-order contributions, the effective spin Hamiltonian

$$\hat{H}' = -J'\left( \sum_{i=1}^{3} \hat{S}'_{i,z}\hat{S}'_{i+1,z} + \hat{S}'_{4,z}\hat{S}'_{1,z} \right) + \mu_B g'_z \hat{S}'_z B_z - \frac{1}{2}\chi'_0 (B_x^2 + B_y^2) \quad (4)$$

is obtained, where $J' = (g'_z / g_z)^2 J = 9J$ and $\chi'_0 = 3\mu_B g_{xy}^2 /(4|D|)$. At low temperatures, the system thus effectively exhibits pure Ising-type interactions. Also, the Zeeman term is then operative only for fields in the z direction, which in conjunction with the Ising interaction produces a level crossing, i.e., a magnetization step in this field direction. Magnetic fields in the plane of the grid, in contrast, are operative only through the second-order term $-1/2\chi'_0(B_x^2 + B_y^2)$ corresponding to magnetic moments, which increase linearly with field (since $m_\alpha = -\partial\langle\hat{H}\rangle/\partial B_\alpha$ at T = 0). Thus, at low temperatures, the [2 × 2] grid **6** behaves like a cluster with pure Ising-type interactions.

Some further points were noted. The linear increase for fields in the xy plane of the grid was found to be directly related to the second-order term in $\hat{H}'$. Its observation thus underscores the importance of second-order contributions in the magnetism of MNMs. Moreover, an effective "amplification" of the magnetic interaction is realized since $J' = 9J$. In fact, the magnetization step was well resolved already at a measurement temperature of 1.9 K only because of this (J' ≈ -13.5 K). And finally, a profound link with the magnetism of metamagnets was pointed out. Metamagnets are characterized by a field-induced transition from the anti- to the ferromagnetic spin configuration (spin-flip) for parallel fields $B > B_{ex}$ ($B_{ex} \propto |J|$ is the exchange field), and a linear increase of the magnetization due to a tilting of the magnetization vectors out of an easy-axis in perpendicular fields [24]. The similarity with the two characteristic features observed in the grid **6** is apparent. Furthermore, metamagnetism appears in extended antiferromagnets when $B_{ex} < B_a$ ($B_a \propto |D|$ is the anisotropy field), mirroring the finding $|J| << |D|$. In fact, Hamiltonian (3) is the finite-size version of a spin Hamiltonian frequently used to discuss metamagnetism at the microscopic level [25]. In the Co(II)-[2 × 2] grid no long-range order develops as intermolecular interactions are negligibly small preventing the system from undergoing a metamagnetic phase transition. However, the analysis of its magnetism can be completely recast in the language of metamagnetism, and in this sense one may speak of a single-molecule metamagnet.

The susceptibility and magnetization curves for the "bromide" and "unsubstituted" Co(II)-grids $[Co_4(\mathbf{4b})_4](PF_6)_8$ (**8**) and $[Co_4(\mathbf{4c})_4](PF_6)_8$ (**9**), respectively, were found to be very similar to those of the "methyl" grid **6** (Fig. 7) [21]. Interestingly, both the maximum in the susceptibility and the field position of the magnetization step increase in the series of the grids

**6**, **8**, and **9**. Since both characteristics are proportionally related to the coupling constant J, this unambiguously evidences a 50% increase of J. The variation is not dramatic, but demonstrates the great potential inherent in these systems of a controlled tuning of the magnetic interactions.

Bis(bipyridyl)-pyrimidine based Fe(II)-[2 × 2] grids displayed another interesting magnetic behavior, which shall be reported here in view of its importance, although it is somewhat beyond the context of this article as it is not covered by a pure spin-Hamiltonian such as Eq. (1). Incorporation of Fe(II) in molecular complexes is interesting as this ion may undergo spin transitions (ST) from a low-spin ($S_i = 0$) to a high-spin ($S_i = 2$) state, which often is hysteretic providing multistability at the molecular level. In the complex $[Fe_4(\mathbf{4d})_4](ClO_4)_8$ (**10**), ST phenomena were indeed observed [26]. $^1$H NMR, magnetic susceptibility, and $^{57}$Fe Mössbauer studies (Fig. 8) demonstrated a gradual multistep ST between three magnetic states with i) three Fe(II) ions in the high-spin (HS) state and one in the low-spin (LS) state, ii) two HS and two LS ions, and iii) one HS and three LS ions. The transition between these states could be triggered by the external perturbations temperature, pressure, and light (light-induced excited spin state trapping, LIESST, is shown in Fig. 8). Extensive investigations on a number of Fe(II)-[2 × 2] grids with modified ligands showed that the ST behavior is tunable over a broad range by the choice of the end-group at the $R_1$ position, and to a lesser extend of those at the periphery [27]. Apparently, the Fe(II)-[2 × 2] grid system is a unique model system for studying multilevel ST arising from communicating metal ions within a cluster.

## 3. Elementary excitations in antiferromagnetic Heisenberg rings

Molecular wheels are distinguished by a virtually perfect ring-like arrangement of the metal ions within a single molecule. The decanuclear wheel $[Fe_{10}(OCH_3)_{20}(O_2CCH_2Cl)_{10}]$, now generally called Fe10, has become the prototype for this class of compounds [28,29], but some octanuclear wheels were previously known [30-32]. In the last decade, many wheels with different metal ions and number of centers in the ring have been realized (only homonuclear wheels with an even number of metal ions, and with antiferromagnetic couplings, are considered here) [33-36].

Antiferromagnetic (AF) wheels were widely regarded as models for one-dimensional (1D) AF chains in view of their structure, implying that physical concepts found for these chains should also describe them. However, the finite-size effects in AF wheels are still strong (*vide infra*), and it turned out that they do not behave like 1D AF chains at all - a point which will be picked up again in section 5. In the current section, the picture of the excitations in AF wheels as it emerged in recent years shall be presented briefly.

A large number of different experiments demonstrated that the magnetic properties of the AF wheels are excellently described by the minimal spin Hamiltonian

$$\hat{H} = -J\left( \sum_{i=1}^{N-1} \hat{\mathbf{S}}_i \cdot \hat{\mathbf{S}}_{i+1} + \hat{\mathbf{S}}_N \cdot \hat{\mathbf{S}}_1 \right) + D \sum_{i=1}^{N} \hat{S}_{i,z}^2 + \mu_B g \hat{\mathbf{S}} \cdot \mathbf{B} , \qquad (5)$$

where N is the number of spin centers with $S_i = s$ for all centers, and g = 2.0. Only four parameters enter Hamiltonian (5): J, D, N, and s. Accordingly, molecular wheels are particularly suited for the kind of physical considerations envisaged in the introduction. In the AF wheels realized so far, the magnetic anisotropy is small and its effects are disregarded in this section (D = 0), which focuses on the energy spectrum and spin dynamics due to the AF Heisenberg interactions. The corresponding model of an AF Heisenberg ring (AFHR) is specified entirely by the two parameters N and s (|J| only sets the energy scale). Only the zero-field situation will be discussed in the following (once the zero-field energy spectrum is known, the effects of a magnetic field are simple to consider [29,37,38]).

The levels of the AFHR may be classified by S and M, the spin quantum numbers belonging to the total spin operator $\hat{\mathbf{S}} = \sum_i \hat{\mathbf{S}}_i$ . The total net magnetic moment is zero in the ground state, corresponding to S = 0. The first phenomenological insight into the structure of the higher lying spectrum came from the observation of steps in magnetization and torque curves at very low temperatures [29,39-42]. Their analysis showed that the lowest states are those with minimal energy for each value of the total spin S = 0, 1, 2, …; and the energies were found to follow closely the Landé rule $E(S) \propto S(S+1)$. In view of the analogy of such an energy level pattern with the spectrum of a rigid rotator, which would be described by the Hamiltonian $\hat{H} = \hat{\mathbf{S}}^2 / 2I$ (I is the moment of inertia), the notion of a rotational mode was introduced for such a set of spin levels [43]. This band of states is also famously known as the "tower of states" in the context of extended AF Heisenberg lattices [44-46]. A subsequent numerical study provided a comprehensive picture of the excitations in AFHRs [47], which is presented here with the example of an octanuclear AFHR of spin-3/2 centers (N = 8, s = 3/2).

The low-lying energies were found to exhibit a remarkable structure when plotted as function of the total spin S (Fig. 9a). Several parallel rotational bands were identified. The lowest band, the L band, which starts from the S = 0 ground state, consists exactly of those states which in the presence of a magnetic field would produce the magnetization/torque steps at low temperatures. The next higher lying rotational bands were denoted collectively as E band. The remaining states were summed up in the so called quasi-continuum (they are actually of little relevance at temperatures $T \lesssim |J|$). This particular structure of the energy spectrum is mirrored by a surprisingly simple structure of the spin-spin correlation functions. Analysis of the spin correlation functions (SCFs) is of high interest because, visualizing directly the spin dynamics, they provide the best view on the elementary excitations in spin systems. Moreover, they are of direct relevance for experimental techniques such as inelastic neutron scattering (INS), nuclear magnetic resonance, or electron paramagnetic resonance.

Two different types of excitations were immediately identified from the SCF (Fig. 9b). At higher frequencies, weakly temperature dependent excitations appear (excitations $E_1$ and $E_2$ in Fig. 9b), reflecting transitions from states of the L band to those of the E band. At lower frequencies, however, another set of excitations is present, which are related to transitions within the L band.

The physical interpretation of these excitations was actually already given more than fifty years ago by Anderson in his AF spin-wave theory [44,45]. In the ground state all the single-ion spins are aligned antiferromagnetically and the spin configuration may be depicted classically by alternating anti-parallel spins. Starting from this ground state configuration, two kinds of excitations are possible. One obvious excitation would be to "flip" one of the spins, or more precisely, to change its magnetic quantum number $m_i$ by one unit, i.e. from $m_i = \pm s$ to $m_i = \pm(s - 1)$ (for a spin with $s > 1/2$ this actually corresponds more to a canting of the spin than to a flip). However, such spin-flip states are not eigenstates of the Heisenberg Hamiltonian, and these excitations accordingly can "hop" along the ring from one site to the next. Eigenfunctions can be constructed by appropriate linear combinations of spin-flip states resulting in delocalized, wave-like excitations - the celebrated spin waves. However, the energy required for a "flip" is on the order of $2|J|$ as one has to break up two AF bonds. In finite AF Heisenberg spin systems there is a second, energetically lower lying, excitation. The orientation of the classical ground-state spin configuration in space is not fixed, allowing for a coherent rotation of all the spins. Quantum mechanically, this rotational degree of freedom corresponds to that of a rigid rotator, and the eigenstates thus belong to $S = 0, 1, 2, \ldots$ . As the underlying coherent motion of the spins may be equivalently described by one single vector, the Néel vector, this excitation mode was named quantized rotation of the Néel vector. The association of these excitations with the structure seen in the energy spectrum and the SCF (Fig. 9) is obvious. The L band and the related low-frequency peaks in the SCF exactly correspond to the quantized rotation of the Néel vector; the E band and the related SCF peaks correspond to spin waves.

The numerical studies and the spin-wave theory produced a detailed list of properties of the excitations which were subjected to experimental test by INS measurements on the chromic wheel $[Cr_8F_8\{O_2CC(CH_3)_3\}_{16}]\cdot 0.25C_6H_{14}$ (**12**), or Cr8 in short [48]. The INS data were successfully fitted to Hamiltonian (5) with J = -16.9 K and D = -0.44 K (s = 3/2, N = 8, B = 0) [49], showing that the magnetic anisotropy is weak and Cr8 indeed is a good model for an AFHR. The experimental INS intensity as function of energy transfer is presented in Fig. 10 for various temperatures. The similarity with the theoretical result for the SCF, Fig. 9b, is apparent. Careful analysis of the INS data allowed one to demonstrate all properties expected from the numerics and spin-wave theory [48].

The above results showed that AF molecular wheels are essentially ordinary spin-wave antiferromagnets – but one important difference has to be noted. In extended antiferromagnets the experimentally observed lowest lying excitations are spin waves. Here, the rotation of the Néel vector becomes unobservable because of its slow dynamics, which undercuts any experimental time scale, i.e. because a Néel-ordered ground state develops. In finite systems such as the molecular wheels, in contrast, the lowest lying excitations correspond to the Néel-vector rotation, and not to spin waves. The AF molecular wheels allow one to observe this fundamental AF excitation mode for the first time – 50 years after its theoretical suggestion.

It is interesting to ask why the above picture of elementary excitations, which is very different from that found in 1D chains, holds for the molecular wheels. It turned out that quantum fluctuations in AFHRs decline with decreasing ring size N and increasing spin length s (for the infinite s = 1/2 chain, quantum fluctuations are strongest) [47]. This gives rise to a diagram as shown in Fig. 11, where the gray area indicates weak quantum fluctuations. Obviously, for the small molecular wheels quantum fluctuations are weak and a (semi)classical theory, such as spin-wave theory, is expected to work, as demonstrated above.

Thus, and this is an important insight, the spin structure due to the Heisenberg interactions is essentially classical in molecular wheels.

This situation allows one to describe the rotation of the Néel vector, or the L band, respectively, as follows. A natural coupling scheme would be to couple first the spins on the AF sublattice A to a total spin $\hat{\mathbf{S}}_A = \sum_{i\in A} \hat{\mathbf{S}}_i$ and those on the sublattice B to $\hat{\mathbf{S}}_B = \sum_{i\in B} \hat{\mathbf{S}}_i$, and then to couple $\hat{\mathbf{S}}_A$ and $\hat{\mathbf{S}}_B$ to yield the total spin $\hat{\mathbf{S}} = \hat{\mathbf{S}}_A + \hat{\mathbf{S}}_B$ (the corresponding wave functions would be written as $|\alpha_A \alpha_B S_A S_B SM\rangle$, where $\alpha_A$ and $\alpha_B$ abbreviate further intermediate quantum numbers). For the lowest states one would expect maximally polarized sublattices, corresponding to $S_A = S_B = Ns/2$. If quantum fluctuations are negligible, then the spin functions $|Ns/2, Ns/2, SM\rangle$ constructed this way for each S = 0, 1, 2, … are essentially the exact eigenfunctions of the L band. As a result, as long as one is concerned only with the states of the L band, one can replace the Heisenberg Hamiltonian $-J\left(\sum \hat{\mathbf{S}}_i \cdot \hat{\mathbf{S}}_{i+1} + \hat{\mathbf{S}}_N \cdot \hat{\mathbf{S}}_1\right)$ by the effective spin Hamiltonian $-a_1 J \hat{\mathbf{S}}_A \cdot \hat{\mathbf{S}}_B$ (the similarity with the Hamiltonian of a rigid rotator should be noted). This is an enormous simplification. In the classical limit, $a_1 = 4/N$, but quantum fluctuations tend to enhance the value of $a_1$ (e.g., for N = 8 and s = 3/2, $a_1$ = 0.5586).

Importantly, this procedure could be used also to derive an effective Hamiltonian, which describes the lowest lying states of the microscopic Hamiltonian (5), including anisotropy and magnetic field [50]. The effective Hamiltonian

$$\hat{H} = -a_1 J \hat{\mathbf{S}}_A \cdot \hat{\mathbf{S}}_B + b_1 D\left(\hat{S}_{A,z}^2 + \hat{S}_{B,z}^2\right) + \mu_B g \hat{\mathbf{S}} \cdot \mathbf{B} \tag{6}$$

was obtained, with $S_A = S_B = Ns/2$. The two parameters were calculated in the classical limit as $a_1 = 4/N$ and $b_1 = (2s-1)/(Ns-1)$, but both are modified as quantum fluctuations become stronger (i.e. the larger the ring size and the smaller the spins become). Comparison of results calculated numerically for the microscopic Hamiltonian (5) and the effective Hamiltonian (6) showed excellent agreement [50]. Hamiltonian (6) is thus a very versatile tool for the analysis of, e.g., the low-temperature magnetization curves of AF molecular wheels.

The picture of the elementary excitations in AF molecular wheels has undergone a drastic change in recent years, eventually emerging as the picture outlined in this section. This may be demonstrated also with the following example. The magnetizations steps were frequently explained qualitatively by arguing that starting from the ground state (with S = 0), the strong applied magnetic field first flips one spin at the first level crossing (leading to a S = 1 state), then a second spin, and so on. However, spin configurations with one flipped spin correspond to spin waves, but these are too high in energy as shown before and thus not responsible for the magnetization steps. In contrary, the magnetization steps originate from the L band, in which no spin flips are involved whatsoever, but represents a coherent rotation of the AF spin configuration.

## 4. Magnetic phenomena in [3 × 3] grids

So far only few [3 × 3] grids incorporating magnetic metal ions have been synthesized. They all are based on the ligand 2POAP (**13a**) and derivatives of it (Fig. 12a). [3 × 3] grids were obtained with the metal ions Mn(II), Fe(III), Co(II), Ni(II), and Cu(II) [51-57]. In the [3 × 3] grid structures, nine metal ions are bridged by six 2POAP ligands so as to form the characteristic 3 × 3 square-matrix like arrangement of the metal centers (Fig. 12b). The molecules generally exhibit (approximate) $D_{2d}$ symmetry with the $S_4$ symmetry axis perpendicular to the grid plane, which corresponds to the uniaxial magnetic axis z. In this section the magnetism of the first two synthesized magnetic [3 × 3] grids, $[Cu_9($**13a**$-H)_6](NO_3)_{12}\cdot 9H_2O$ (**14**) and $[Mn_9($**13a**$-2H)_6](ClO_4)_6\cdot 3.75CH_3CN\cdot 11H_2O$ (**15**) [51,52], is reviewed with emphasis on the Mn(II)-[3 × 3] grid. The situation for the other synthesized magnetic [3 × 3] grids is covered only briefly.

For an idealized [3 × 3] structure, the microscopic Hamiltonian (1) results in the spin Hamiltonian

$$\hat{H} = -J_R\left(\sum_{i=1}^{7}\hat{\mathbf{S}}_i\cdot\hat{\mathbf{S}}_{i+1}+\hat{\mathbf{S}}_8\cdot\hat{\mathbf{S}}_1\right)-J_C\left(\hat{\mathbf{S}}_2+\hat{\mathbf{S}}_4+\hat{\mathbf{S}}_6+\hat{\mathbf{S}}_8\right)\cdot\hat{\mathbf{S}}_9+D_R\sum_{i=1}^{8}\hat{S}_{i,z}^2+D_C\hat{S}_{9,z}^2+\mu_B g\hat{\mathbf{S}}\cdot\mathbf{B}, \quad (7)$$

which generally describes the magnetism of [3 × 3] grids well (for the numbering of the spin centers see Fig. 12b). $J_R$ characterizes the nearest-neighbor couplings within the eight peripheral metal ions, and $J_C$ those involving the central metal ion. $D_R$ and $D_C$ are the anisotropy constants of the peripheral and central metal ions, respectively. The $D_{2d}$ symmetry would allow in principle for different anisotropies, $D_{Rc}$ and $D_{Re}$, of the corner and edge ions, respectively. This difference, however, has little effect on the low-lying energy spectrum and the combined effect of $D_{Rc}$ and $D_{Re}$ is well described by one parameter $D_R$.

For the Cu(II)-[3 × 3] grid **14**, interesting aspects emerged concerning the exchange coupling situation and the origin of magnetic anisotropy [58]. The magnetic susceptibility as a function of temperature demonstrated the simultaneous presence of both antiferromagnetic and ferromagnetic couplings in the grid (Fig. 13). The data fit well to Hamiltonian (7) with $J_R$ = 0.75 K, $J_C$ = -35 K, g = 2.3 ($S_i$ = 1/2, $D_R = D_C = 0$). The magnetization curve at low temperatures indicated a S = 7/2 ground state. The spin structure in the ground state may thus be rationalized by having all the spins on the periphery pointing up and the central spin pointing down. The observation of antiferromagnetic couplings to the central Cu ion was noted to be inconsistent with a $d_{z^2}$ magnetic orbital of this ion inferred from the crystal structure. A fluxonial state of the central Cu ion was suggested [51,54,58].

The limit $|J_R| << |J_C|$ allowed a perturbational rationale of the energy spectrum. For $J_R = 0$, the Cu grid decomposes into a pentanuclear "star" with antiferromagnetic couplings (consisting of the centers 2, 4, 6, 8, and 9), and four "free" Cu ions. The energy spectrum thus exhibits the level pattern of a pentanuclear star, but with an additional weak splitting of the levels due to the ferromagnetic exchange $J_R$ (Fig. 14). In particular, the lowest energy level is split into four S = 1/2, six S = 3/2, four S = 5/2, and the S = 7/2 ground state level.

High-field torque measurements at low temperatures demonstrated the presence of significant magnetic anisotropy in the Cu(II)-[3 × 3] grid **14** [58]. The investigation of magnetic anisotropy in exchange-coupled clusters of spin-1/2 ions, such as Cu(II), is very interesting as these ions do not exhibit a single-ion zero-field splitting [i.e., the term $\sum \mathbf{S}_i \cdot \mathbf{D}_i \cdot \mathbf{S}_i$ in Hamiltonian (1), or the $D_R$- and $D_C$-term in Hamiltonian (7), produce no effect and can be set to zero]. Therefore, in spin-1/2 clusters the zero-field splitting of spin levels is due to anisotropic exchange and dipole-dipole interactions only (antisymmetric exchange can be usually disregarded). The effect of dipole-dipole interactions can be calculated reasonably well. Anisotropy studies on spin-1/2 clusters thus provide exceptional insight into the origin of anisotropic exchange. Such a study revealed an anisotropy of the antiferromagnetic exchange couplings in the Cu(II)-[3 × 3] grid. Employing a novel experimental scheme, thermodynamic spectroscopy, which exploited the fact that in strong magnetic fields only the S = 7/2 level is populated at low temperatures (Fig. 14), the exchange anisotropy was quantified as $D_{ex} = 0.11(3)$ K, in reasonable agreement with the theoretical expectation $D_{ex} \approx (\Delta g/g)^2 J_C = 0.13$ K ($\Delta g = g_z - g_{xy}$ was determined experimentally to be $\Delta g = -0.14$). A further result of the torque studies, which shall only be mentioned here, was to demonstrate the importance of usually neglected higher-order terms [58].

In the Mn(II)-[3 × 3] grid **15**, antiferromagnetic exchange couplings on the order of a few Kelvins and a S = 5/2 ground state were inferred from magnetic susceptibility and magnetization curves [59]. High-field torque measurements at low temperatures (T = 1.75 K) demonstrated an unusual field dependence of the magnetic anisotropy of the cluster: as a function of field, the anisotropy changes its sign at a critical field of ≈ 7.5 T, from easy-axis to hard-axis type. Subsequent more detailed measurements, at much lower temperatures (T = 0.4 K), revealed spectacular quantum magneto-oscillations in the field dependence of the torque signal (Fig. 15) [60]. This novel effect was investigated in further detail, in particular its dependence on the angle of the applied magnetic field, and a theoretical model was devised which described the data excellently (Fig. 16). This model and the underlying physics will be the following subject of interest.

The analysis of magnetic data for a Mn(II)-[3 × 3] grid is non-trivial because of the huge Hilbert space. For a cluster of nine spin-5/2 centers it consists of 10 077 696 states. Even in the simplest case of retaining only isotropic exchange terms in the spin Hamiltonian, and exploiting all symmetries in the reduction of the Hamilton matrix (spin rotational and spatial $D_4$ symmetry), the largest matrix remains of dimension 22 210. Accordingly, an accurate determination of the coupling constants from the susceptibility data has not yet been achieved. However, it was noted that the magnetic susceptibility can be adequately reproduced by considering the [3 × 3] grid structure as an octanuclear "ring" with a metal ion embedded in its center (the ring is formed by the peripheral metal ions) [59]. This viewpoint turned out to be very useful as it enables advantage to be taken of the insights gained from the AFHRs.

The analogy of the Mn(II)-[3 × 3] grid with a doped ring was further supported by the analysis of inelastic neutron scattering measurements and extensive numerical work [61,62]. The INS measurements were performed on a non-deuterated powder sample of the complex $[Mn_9($**13a**$-2H)_6](NO_3)_6 \cdot H_2O$ (**16**), which is a close analogue of the Mn(II)-[3 × 3] grid **15**. Excellent data was obtained (Fig. 17) and with a sophisticated analysis the coupling and anisotropy constants were estimated to be $J_R = J_C = -5.0$ K and $D_R = D_C = -0.14$ K, respectively (concerning the latter values it should be recalled that here they include the effects of the dipole-dipole interactions, in contrast to Ref. [61], see [10]). The INS transitions at high energies evidenced a small but significant deviation of the exchange-coupling situation from ideal $D_{2d}$ symmetry, in accordance with the crystal structure of the cluster, but we shall not dwell on this detail here. The low-lying energy spectrum due to the isotropic exchange terms was calculated [61], and is presented in Fig. 18 as function of the total spin S. The similarity of the energy-level pattern with that of an AFHR is striking. In particular, an L band, an E band, and a quasi continuum can be identified also in the Mn(II)-[3 × 3] grid, and the physical interpretation of these excitations is analogous to that for the AFHR. The main difference here is that the L band starts with a S = 5/2 state.

As in the case of an AFHR, only the states of the L band are of relevance for the field dependent low-temperature properties. Accordingly, the recipe used for the AFHR to construct an effective spin Hamiltonian for describing the L band, see section 3, was applied to the Mn(II)-[3 × 3] grid. This led to the effective Hamiltonian

$$\hat{H} = -\tilde{J}_R \hat{\mathbf{S}}_A \cdot \hat{\mathbf{S}}_B - J_C \hat{\mathbf{S}}_B \cdot \hat{\mathbf{S}}_9 + \tilde{D}_R \left( \hat{S}_{A,z}^2 + \hat{S}_{B,z}^2 \right) + D_C \hat{S}_{9,z}^2 + \mu_B g \hat{\mathbf{S}} \cdot \mathbf{B} \, . \tag{8}$$

where $\tilde{J}_R = 0.526\, J_R$ and $\tilde{D}_R = 0.197\, D_R$ for the Mn(II)-[3 × 3] grid [60,62]. The corner and edge spins were coupled such as to yield the sublattice spins $\hat{\mathbf{S}}_A = \sum_{i=1,3,5,7} \hat{\mathbf{S}}_i$ and $\hat{\mathbf{S}}_B = \sum_{i=2,4,6,8} \hat{\mathbf{S}}_i$, respectively, with $S_A = S_B = 10$. A detailed comparison with numerical results for the microscopic Hamiltonian (7) demonstrated that the effective Hamiltonian (8) works remarkably well for a wide range of $J_C/|J_R|$ values [62]. $J_C$ in fact may assume any value ($J_R$ is of course limited to $J_R < 0$). The good performance of Hamiltonian (8) is also underpinned by the good agreement of the observed and calculated quantum magneto-oscillations shown in Fig. 16. Apparently, Hamiltonian (8) grasps the essential physics. The reason for this is that the spin structure due to the dominant Heisenberg interactions is again essentially classical in the Mn(II)-[3 × 3] grid.

The quantum magneto-oscillations in the field dependence of the torque shall now be explained briefly [62]. In an applied magnetic field, the Zeeman splitting of the spin levels S = 5/2, 7/2, 9/2, … leads to a series of ground-state level crossings at characteristic fields, where the ground state switches from S = 5/2 to S = 7/2, S = 7/2 to S = 9/2, and so on (Fig. 19a). In the absence of magnetic anisotropy this would result in a staircase-like magnetization curve at low temperatures, as for example is familiar for the molecular wheels. However, in Mn(II)-[3 × 3] the magnetic anisotropy produces a strong mixing of the spin levels at the level-crossing fields, leading to avoided level crossings (see Fig. 19a). The magnetization curve is not much affected by this, it still shows the typical staircase behavior but with more broadened steps. The torque curve, however, shows a strikingly different behavior. It can be shown that the change in the torque signal at an avoided level crossing is due to two contributions [62], one which is proportional to the magnetization (and accordingly leads to broadened staircases again), and secondly, one which produces a sharp peak in the torque signal centered at the level-crossing field. In Mn(II)-[3 × 3], the first contribution is outweighed by the second, and the torque consists of a series of peaks at the level-crossing fields which superimpose to give the observed oscillatory field dependence (Fig. 19b).

The torque peaks arise only for avoided level crossings and are thus a direct signature of level mixing. In other words, the quantum magneto-oscillations in the torque are a striking example of an effect which is not covered by the strong exchange limit usually used to interpret the magnetism of molecular nanomagnets with dominating Heisenberg interactions. Another point is worth mentioning. The intimate connection of the quantum magneto-oscillations and the presence of an L band is apparent from the above discussion. Thus, their observation establishes the first experimental demonstration of such a band in square-matrix like Heisenberg systems. This is of great relevance in the context of two-dimensional antiferromagnetic Heisenberg spin systems; a point which will be picked up again in section 5.

Based on the ligand 2POAP, **13a**, which can be derivatized at several positions (Fig. 12a), a zoo of ligands **13b** – **13h** was synthesized [53-56]. With the exception of **13c**, all ligands gave Mn(II)-[3 × 3] grid complexes with structures very similar to that of the parent grid **15** [53,55,56]. The magnetic susceptibilities, which were measured for most of these complexes, also showed similar behavior as **15**, with slightly different coupling constants (the values for $J_R$ ranged from -5 K to -6.8 K, $J_C$ could not be determined and was assumed to be zero). Besides **14**, further Cu(II)-[3 × 3] grid complexes were obtained using ligands **13a**, **13b**, and **13c** (in some cases the grid complex contained an additional ligand) [52,54,57]. The magnetism was found to be very similar to the parent grid **14**. A Ni(II)-[3 × 3] and Co(II)-[3 × 3] grid were synthesized using ligand **13e** and **13a**, respectively [53]. In both complexes, magnetic susceptibility curves indicated antiferromagnetic exchange interactions. For the Ni(II)-[3 × 3] complex, $J_R = J_C = -17.4$ K and $g = 2.32$ were obtained ($S_i = 1$, $D_R = D_C = 0$). An interesting magnetic behavior was reported for the Fe(III)-[3 × 3] grid $[Fe_9($**13a**$-2H)_6](NO_3)_{15}\cdot 18H_2O$ (**17**) [53,55]. The magnetic susceptibility curve, Fig. 20a, shows first a slight drop followed by a steep rise with decreasing temperatures, very similar to that found in the Cu(II)-[3 × 3] grids. From this profile the simultaneous presence of pronounced antiferromagnetic and weaker ferromagnetic exchange interactions was concluded, but further analysis was not attempted. The magnetization curve, Fig. 20b, suggested a high-spin ground state with $S \approx 29/2$. Mössbauer studies provided evidence for nine high-spin Fe(III) ions in the grid, but the spectra exhibited a pronounced temperature dependence, which was interpreted as a possible indication of slow relaxation of the magnetization at low temperatures [55].

**5. Conclusions and outlook**

The above sections demonstrate that in the [2 × 2] and [3 × 3] grid molecules rarely observed, novel, or sometimes even spectacular magnetic phenomena can be observed. The high molecular symmetry of the clusters - resulting in models describing the magnetic properties with a minimal number of magnetic parameters - and careful studies of the magnetic anisotropy were instrumental in the process of identifying these phenomena. Since the appearance of the single-molecule magnets, understanding, and eventually controlling magnetic anisotropy and its interplay with exchange interactions has become a major issue in this field. The grid molecules provided some beautiful examples in this regard, sometimes of textbook quality, enlightening some aspects of the interplay between anisotropy and Heisenberg exchange.

The experimental and numerical/theoretical studies on the molecular wheels unfolded a transparent picture of the elementary excitations in antiferromagnetic Heisenberg rings, from which the useful concept of a "classical spin structure" emerged. This concept comprises a rotational band structure of the energy spectrum and a recipe for the construction of a low-temperature effective spin Hamiltonian. Importantly, it seems to be relevant not only for the molecular wheels, but for a broader class of antiferromagnetic Heisenberg spin systems (AFHSS), with the Mn(II)-[3 × 3] grid as a striking example. Carrying over the ideas developed for the wheels enabled a comprehensive understanding of the magnetism in this grid, which otherwise would have been difficult to gain in view of the huge Hilbert space of the system.

Further support for the generality of these ideas has been obtained from numerical calculation of the energy spectra for a number of different AFHSS. The L band, or the analogue to it, was found in rings with an odd number of centers, in polytopes such as the tetrahedron, cube, octahedron, icosahedron, triangular prism, axially truncated icosahedron, icosidodecahedron, and in finite triangular lattices [43,63-67,46]. Moreover, the analogue of the E band could also be found, and even the spin-correlation functions showed amazingly similar profiles [47]. For instance, the spin-correlation function of a hexanuclear ring and an octahedron basically coincide, despite the very different coupling topologies of these two kinds of AFHSS [47]. Thus, there is growing evidence that the findings for the even rings are characteristic for a larger class of AFHSS.

The Mn12 molecule represents another interesting example. The exchange coupling topology in Mn12 is characterized by four exchange constants (Fig. 21a). Several numerical studies consistently concluded that the interactions $J_1$ and $J_2$ (both antiferromagnetic) are dominant, while the $J_3$ and $J_4$ interaction paths are weak [68,69]. It is then possible to morph the exchange graph into the one shown in Fig. 21b, unraveling an analogy with the rings. Thus, disregarding the $J_3$ and $J_4$ interactions, a classical spin structure is also expected in Mn12 (as each of the pheripheral Mn ions interact only with one Mn ion in the ring, their coupling does not disturb the classical spin structure in the ring). The interactions $J_3$ and $J_4$ introduce some frustration effects, but being weak they are not expected to be critical. Such considerations actually led to the prediction of a classical spin structure in Mn12 [48], which has been confirmed nicely by recent numerical work [69]. This should be regarded as a remarkable success of the "classical spin structure" concept. In some sense the standard procedure of assigning up and down spins to each Mn center for explaining the $S = 10$ ground state in Mn12 - a consideration which is intrinsically classical - received a belated justification.

However intuitive and satisfying the concepts have appeared in this work, some important issues need to be addressed. One is the issue of dimensionality. From a strict thermodynamical point of view, molecular nanomagnets are of course zero-dimensional (0D) objects, as they do not exhibit long-range effects. However, it seems perfectly obvious to be somewhat less strict here. For instance, it makes sense to consider a single-molecule magnet as a "0D" object, since magnetically it can be treated as a single-spin, i.e., a point-like object. However for the molecular wheels, which exhibit S = 0 ground states, such an approach would not be very useful as the interesting aspects would be completely missed. A more sophisticated approach, which accounts for the ring-like topology of the clusters, was needed. It is thus natural to consider a wheel as a "1D" object; and a [N × N] grid as a "2D" one (for N > 2). However, it has become clear that the molecular wheels do not show any traces of the behavior typical of 1D Heisenberg chains. On the contrary, their characteristic features were well explained by spin-wave theory, because of the weak quantum fluctuations in the wheels (see Fig. 11). At this point, it is interesting to note that spin-wave theory is known to work well in 2D and 3D situations, but to have some serious drawbacks in 1D. The reason again lies in the quantum fluctuations, which are weak in 2D and 3D lattices and strong in the 1D case. Accordingly, the 0D AFHSS have more in common with 2D and 3D systems than with the 1D chain.

Unfortunately, there is no satisfying mathematical apparatus available so far for the consistent treatment of "classical" AFHSS. For instance, the author is not aware of any analytical method, which would allow the calculation of the energies of the E band. With standard spin-wave theory one may calculate for example the energies of the S = 1 states in the case of rings, but not those of the S = 2, 3, ... states (see Fig. 9a). For extended systems this turns out to be sufficient, but not for finite ones. Spin-wave theory has been powerful enough to suggest a clear picture of the elementary excitations in finite AFHSS, if quantum fluctuations are weak, but by itself does not allow consistent calculations.

There has been some progress in the recent decades. Standard spin-wave theory was extended into the so-called modified spin-wave theories [70,71] and finite-size spin-wave theories [72,73] for example. The latter ones are particularly appealing in the present context, as they almost trivially produce the L band. Concerning the E band, however, these methods are plagued by similar obstacles as the ordinary spin-wave theory.

It is interesting to have a closer look at the motivation behind the finite-size spin-wave theories. The discovery of the high-temperature superconductors, which are characterized by 2D antiferromagnetic $CuO_2$ double-layers, initiated a huge interest in the properties of 2D antiferromagnetic Heisenberg lattices [74]. The obviously important question, when does long-range antiferromagnetic order (LRO) exists in such lattices (at T = 0), has not yet been fully answered. Numerical studies would be helpful, but the energy spectrum can be calculated only for small clusters, for which finite size-effects are pronounced. Reliable methods for the finite-size scaling of numerical results are thus needed, and finite-size spin-wave theory aims at providing them [72]. The existence of LRO can in principle be easily seen by an inspection of the lowest lying levels in the spectrum [66,46]. As a matter of fact, LRO in the infinite 2D lattice correlates to a set of states in the finite cluster, called the quasi degenerate joint states (QDJS), or "tower of states" [66,46], whose energies increase with the total spin S according to $E(S) \propto S(S+1)$ - exactly as in the L band (at this point the notations QDJS, tower of states, and L band turn out to be tautologies). This puts the observation of the L band via the quantum magneto-oscillations in the Mn(II)-[3 × 3] grid in another perspective; it amounts to the first clear experimental proof of these levels. The existence of a deep link between the level structure in grids and the question of LRO in 2D Heisenberg lattices is surely surprising, and gratifying. The synthesis of larger magnetic grids would be highly desirable; first progress in this direction has been reported [75].

A final comment shall be added. The existing methods, some of them have been mentioned above, do not provide any criterion as to decide when an AFHSS behaves classically, or when quantum fluctuations are weak. The dependence of the situation on, for example, the values of the exchange constants is subtle. Numerical calculations indicated that in some cases the rotational band structure is rather robust against variations of the coupling constants, while in other cases it is not [76]. A criterion which would allow one to single out, just from the mere knowledge of the coupling matrix $J_{ij}$ and the vector $S_i$, which AFHSS fall into the class of being "classical", would be a major step forward in the kind of game considered in the introduction. This review aimed at showing that many pieces of the puzzle are known - but also that they have not yet fallen into place. It hopefully will provide some guide in where to go.

**Acknowledgements**

It is a pleasure to seize this opportunity and to thank several individuals whose contribution to this work was essential in one way or the other. In particular, special thanks go to Laurie Thompson and his group, to Jean-Marie Lehn and especially Mario Ruben, to Roberto Caciuffo and coworkers, especially Tatjana Guidi and Stefano Carretta, and finally, to my colleagues here in Bern, Roland Bircher, Chris Dobe, Stefan Ochsenbein, Ralph Schenker, Andreas Sieber, and in particular to Hans Güdel. Stefan Ochsenbein and Chris Dobe are gratefully acknowledged for having read the manuscript carefully. Over the years, this work was financially supported by the Deutsche Forschungsgemeinschaft, the Department of Energy (Grant No. DE-FG02-86ER45271), the TMR Program of the European Community, the EC-RTN-QUEMOLNA (Contract n° MRTN-CT-2003-504880), and the Swiss National Science Foundation, which is gratefully acknowledged.

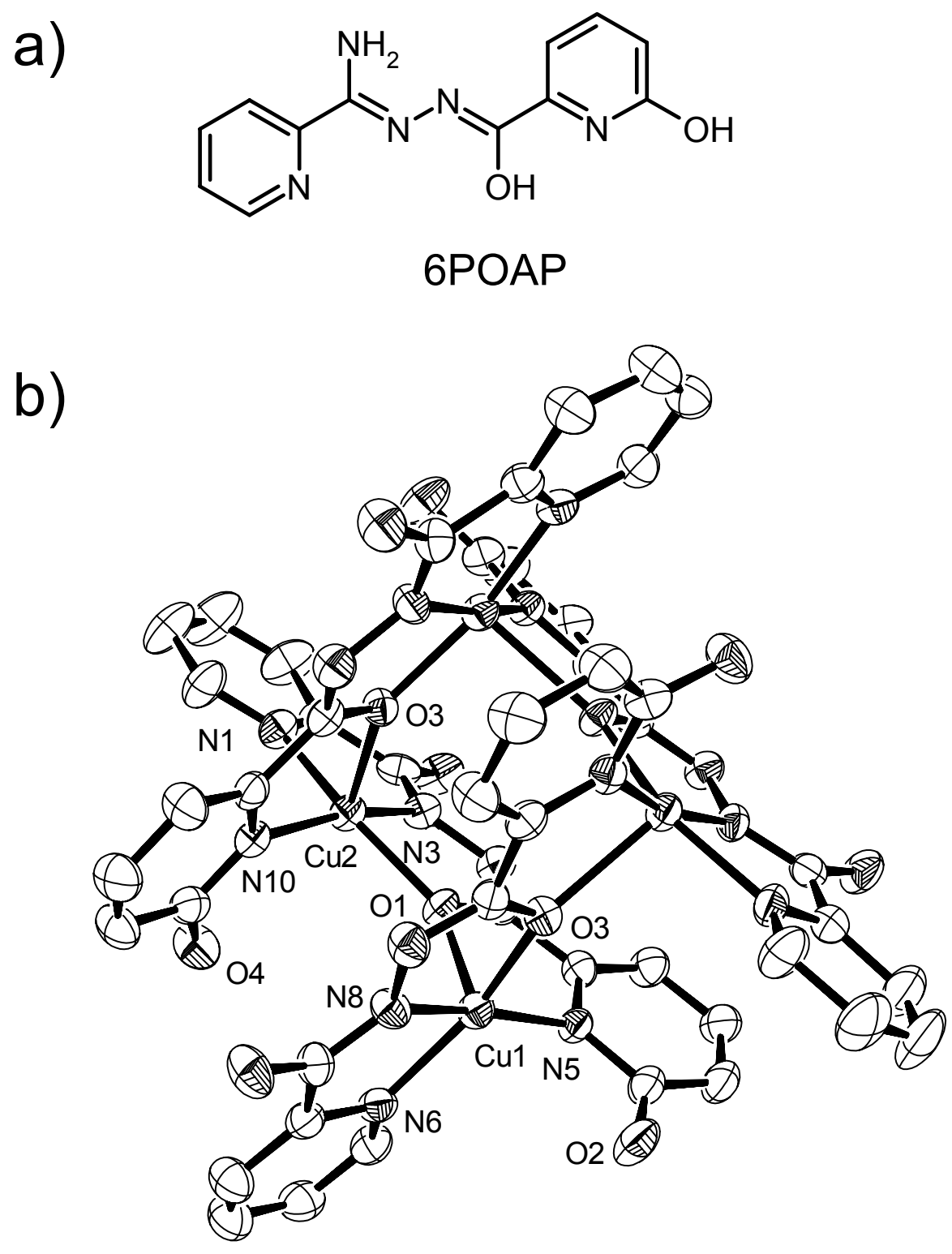

Fig. 1. (a) The ligand 6POAP and (b) structural representation of the Cu(II)-[2 × 2] grid $[Cu_4(6POAP\text{-}H)_4]^{4+}$ (**1**) (40% probability thermal ellipsoids, H atoms omitted).

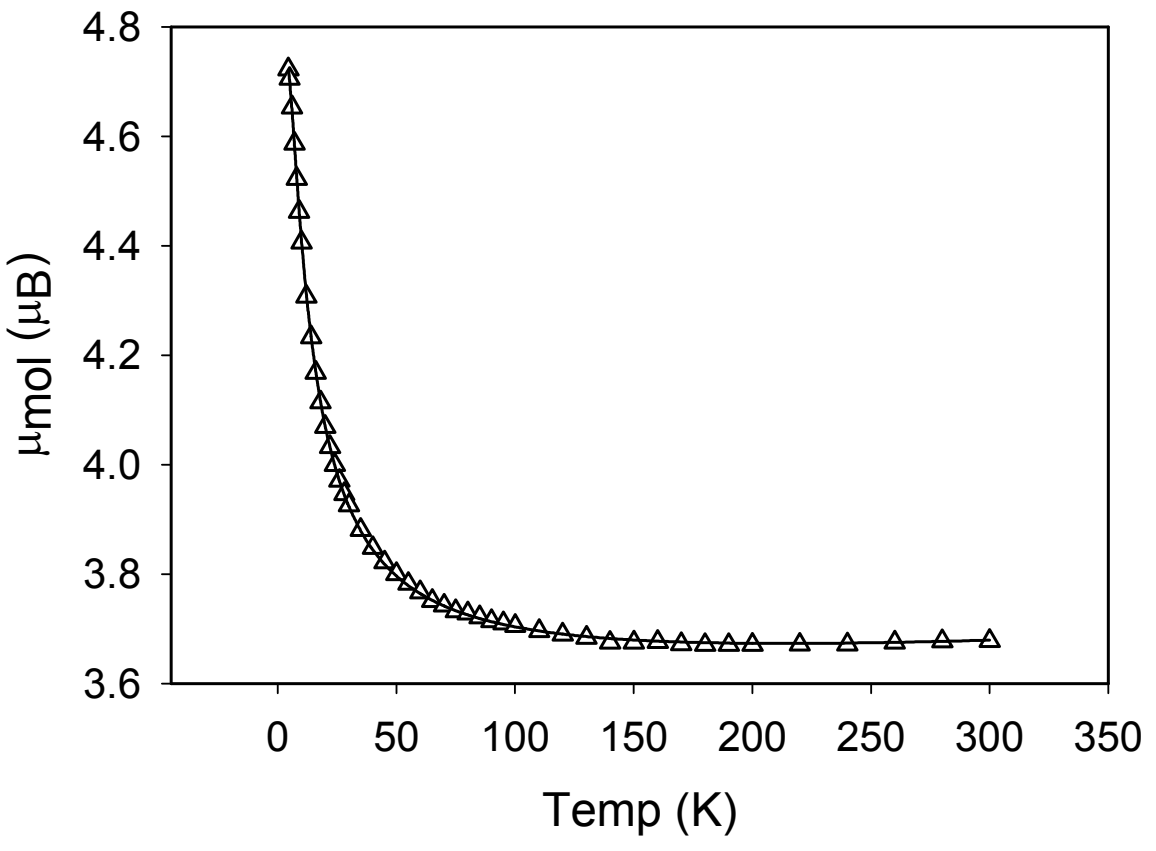

Fig. 2. Temperature dependence of the effective magnetic moment for the Cu(II)-[2 × 2] grid $[Cu_4(6POAP\text{-}H)_4](ClO_4)_4$ (**1**). The solid line represents a fit to the data using Hamiltonian (3), yielding g = 2.060(7), J = 27(1) K (D = Δg = 0). Impurities, $\rho = 4\times10^{-5}$, intermolecular interactions, θ = -0.4 K, and TIP = $232\times10^{-6}$ emu/mol were additionally taken into account. Adapted from ref. [14].

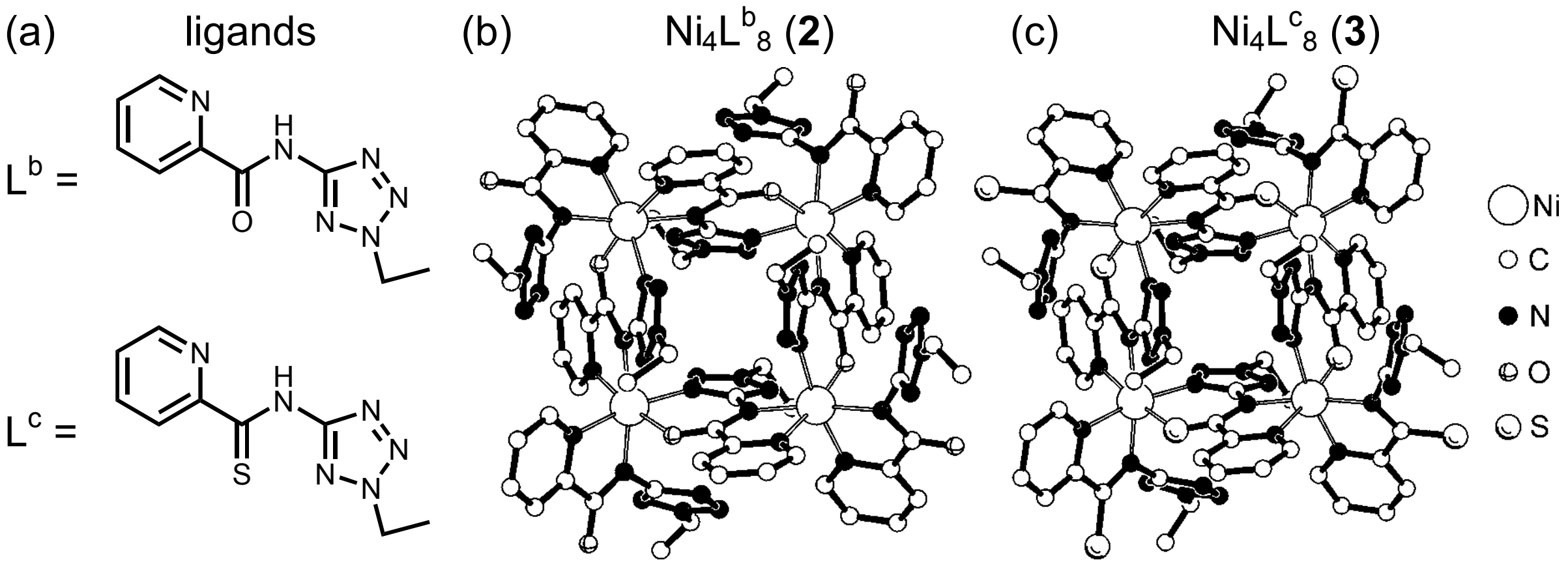

Fig. 3. (a) Sketch of the ligands $L^b$ and $L^c$, and structural representation of the Ni(II) squares (b) $[Ni_4L^b{}_8]$ (**2**) and (c) $[Ni_4L^c{}_8]$ (**3**) viewed along the crystallographic $S_4$ axis (H atoms omitted). Adapted from ref. [18].

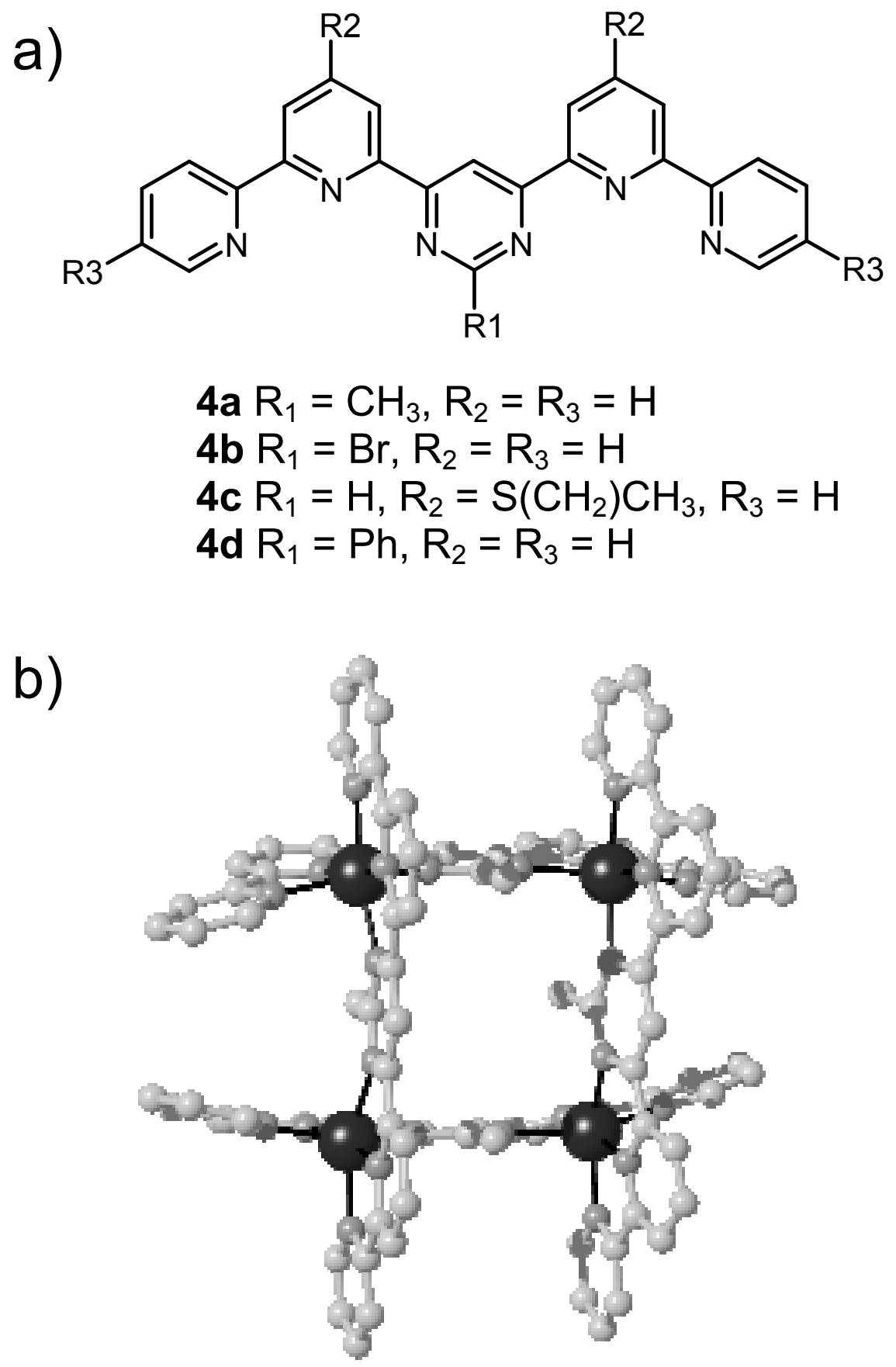

Fig. 4. (a) Bis(bipyridyl)-pyrimidine ligand and (b) crystal structure of the Co(II)-[2 × 2] grid $[Co_4(\mathbf{4a})_4]^{8+}$ (**6**) (H atoms are omitted).

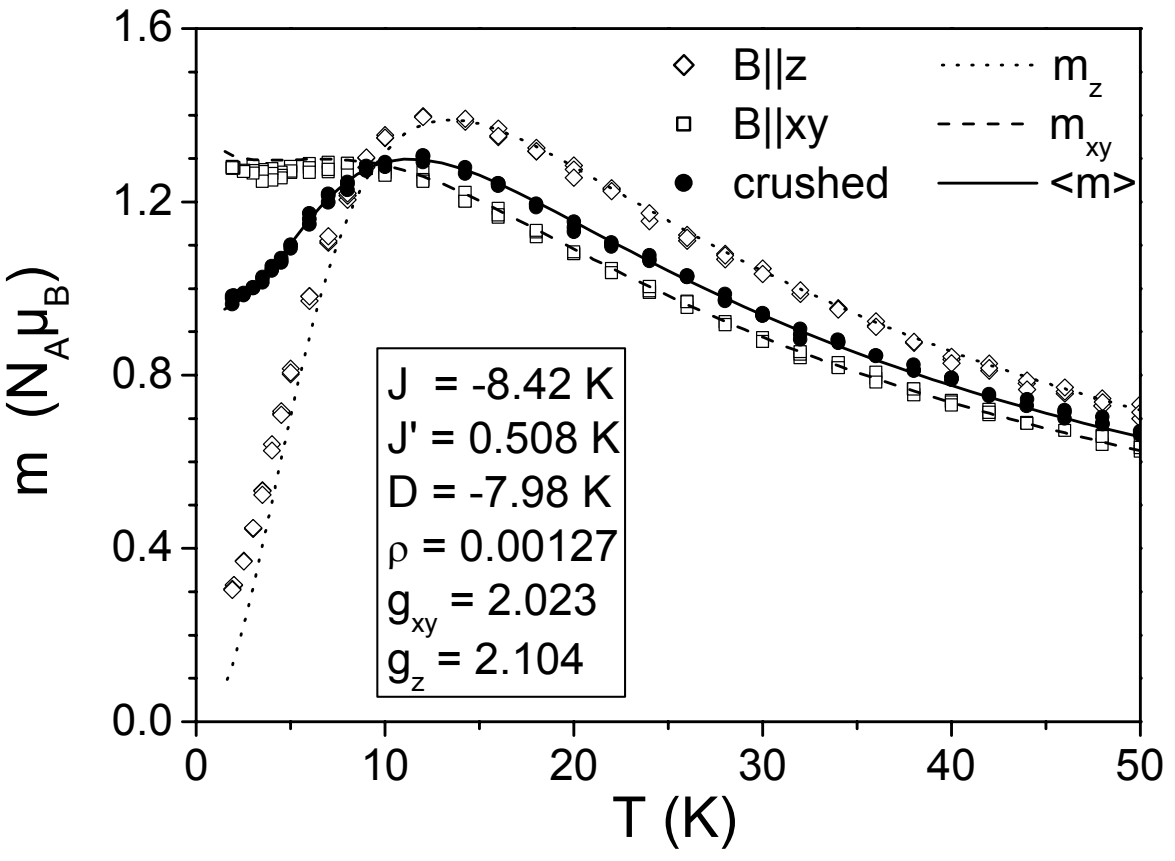

Fig. 5. Temperature dependence of the magnetic moment of single crystals of the Ni(II)-[2 × 2] grid complex $[Ni_4(\mathbf{4a})_4](PF_6)_8$ (**5**) with a magnetic field of 5.5 T parallel to the z axis (open diamonds) and to the xy plane (open squares), respectively, and for a sample of crushed crystals (solid circles). Lines represent fits using the Hamiltonian (3) extended by biquadratic exchange (and a spin-1 impurity with concentration ρ) with parameters given in the panel. Adapted from ref. [20].

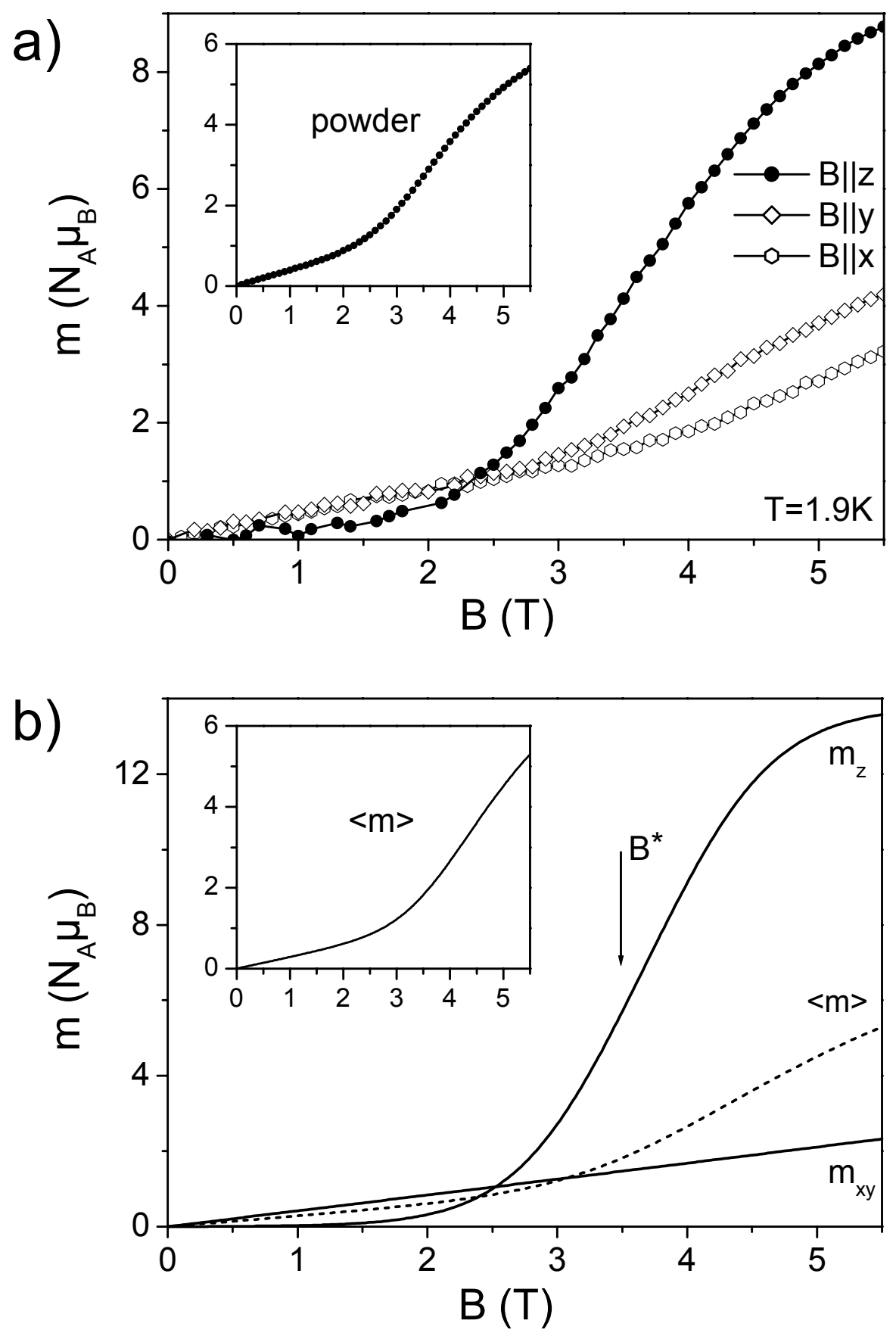

Fig. 6. (a) Field dependence of the magnetic moment of a single crystal of the Co(II)-[2 × 2] grid complex $[Co_4(\mathbf{4a})_4](PF_6)_8$ (**6**) at 1.9 K for magnetic fields along the main axes. The inset displays the magnetization curve of a powder sample of **6**. (b) Field dependence of the magnetic moment as calculated with Hamiltonian (3) for J = -1.8 K, D = -20 K, and g = 2.3 at 1.9 K for fields in the z direction and the xy plane. The dashed line and the inset show the averaged magnetic moment corresponding to powder samples. Adapted from ref. [21].

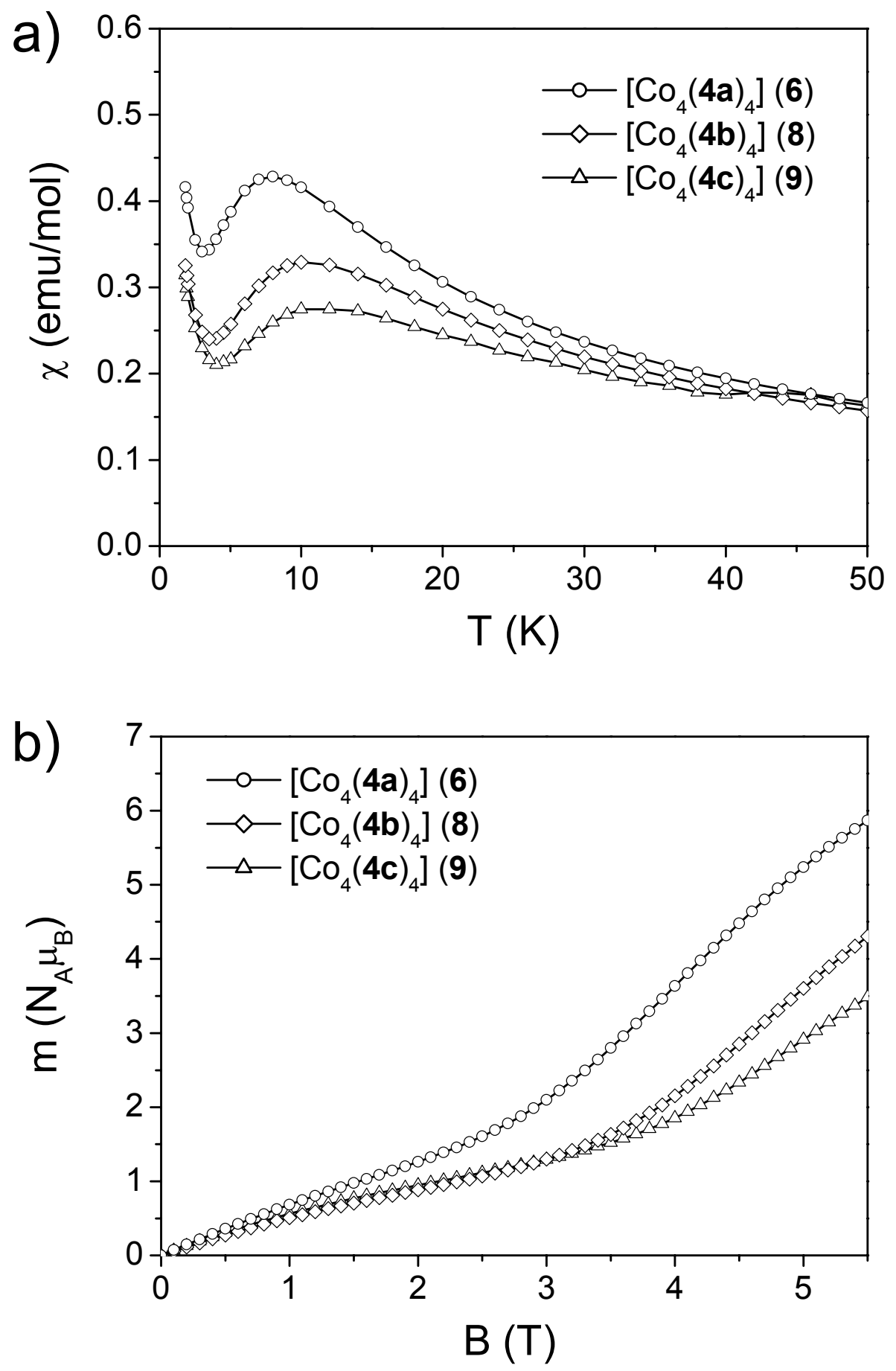

Fig. 7. (a) Magnetic susceptibility vs. temperature and (b) magnetic moment vs. field at 1.9 K for powder samples of the Co(II)-[2 × 2] grids **6**, **8**, and **9**. Adapted from ref. [21].

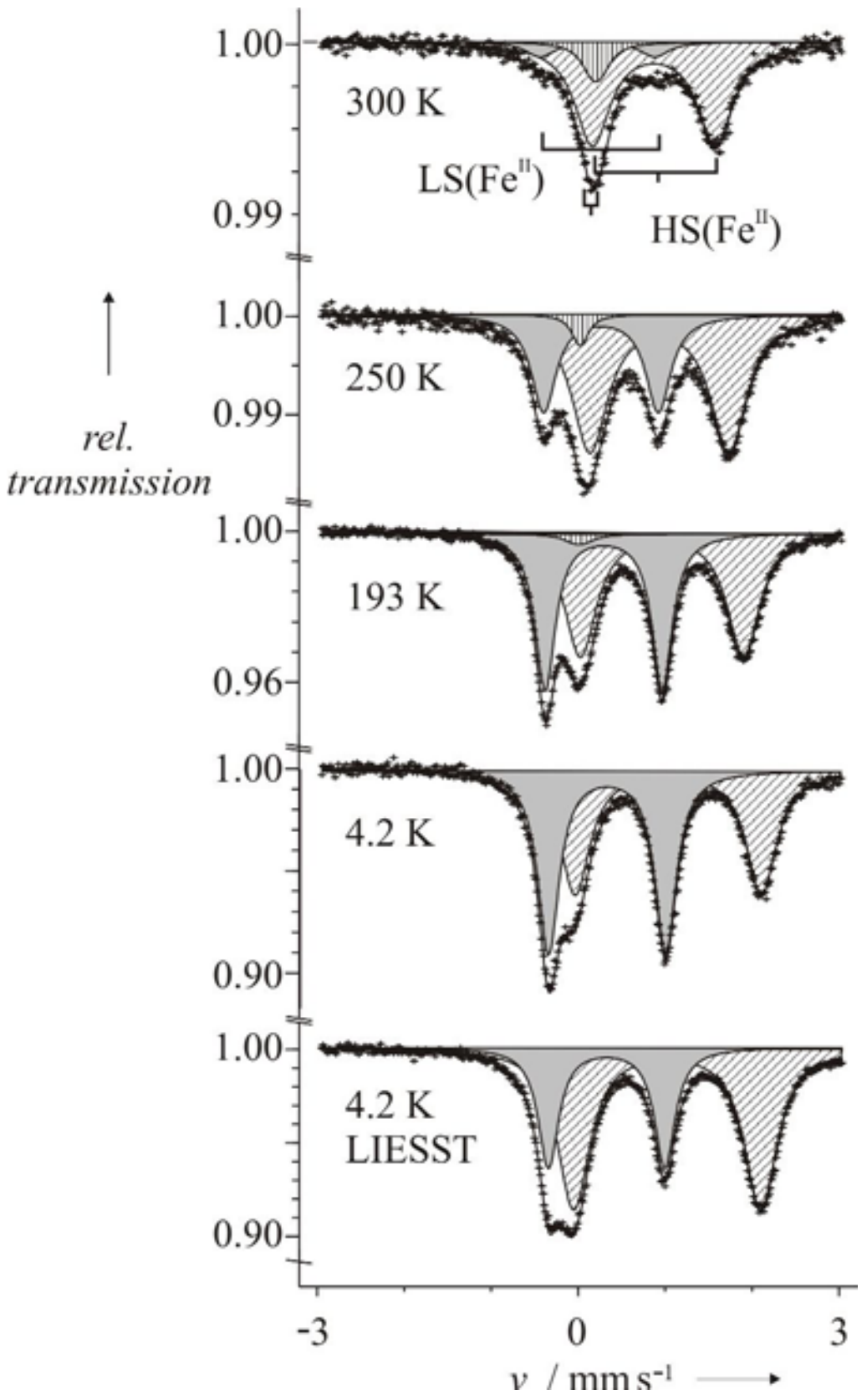

Fig. 8. [57]Fe Mössbauer spectra for the Fe(II)-[2 × 2] grid complex $[Fe_4(\mathbf{4d})_4](ClO_4)_8$ (**10**) showing the increase of the low-spin fraction towards lower temperatures. The lowest panel shows the 4.2 K spectra as recorded after irradiation with λ = 514 nm light (LIESST effect). Reproduced with permission of the copyright holders from ref. [26].

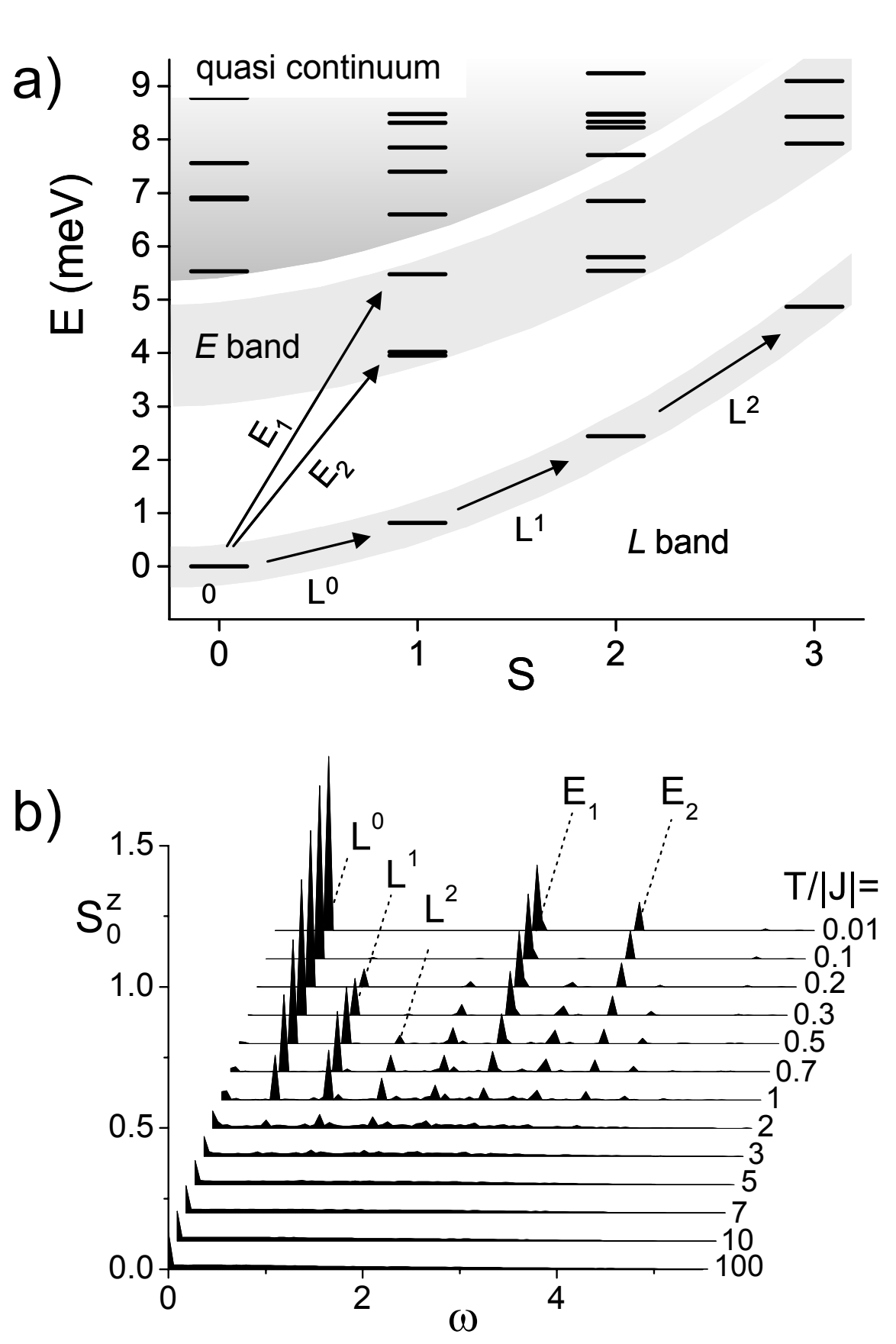

Fig. 9. Energy spectrum and spin autocorrelation function of an octanuclear spin-3/2 antiferromagnetic Heisenberg ring. (a) Energy spectrum vs. total spin quantum number S. The coupling constant was set to J = -16.9 K as appropriate for the molecular wheel Cr8 (**12**). Arrows indicate observed transitions and their labeling. (b) Normalized spin pair-autocorrelation function $S_0^z(T,\omega)$. Only that part of the spectrum with nonzero peak amplitudes is shown. Peaks were labeled as in panel (a). Adapted from ref. [47].

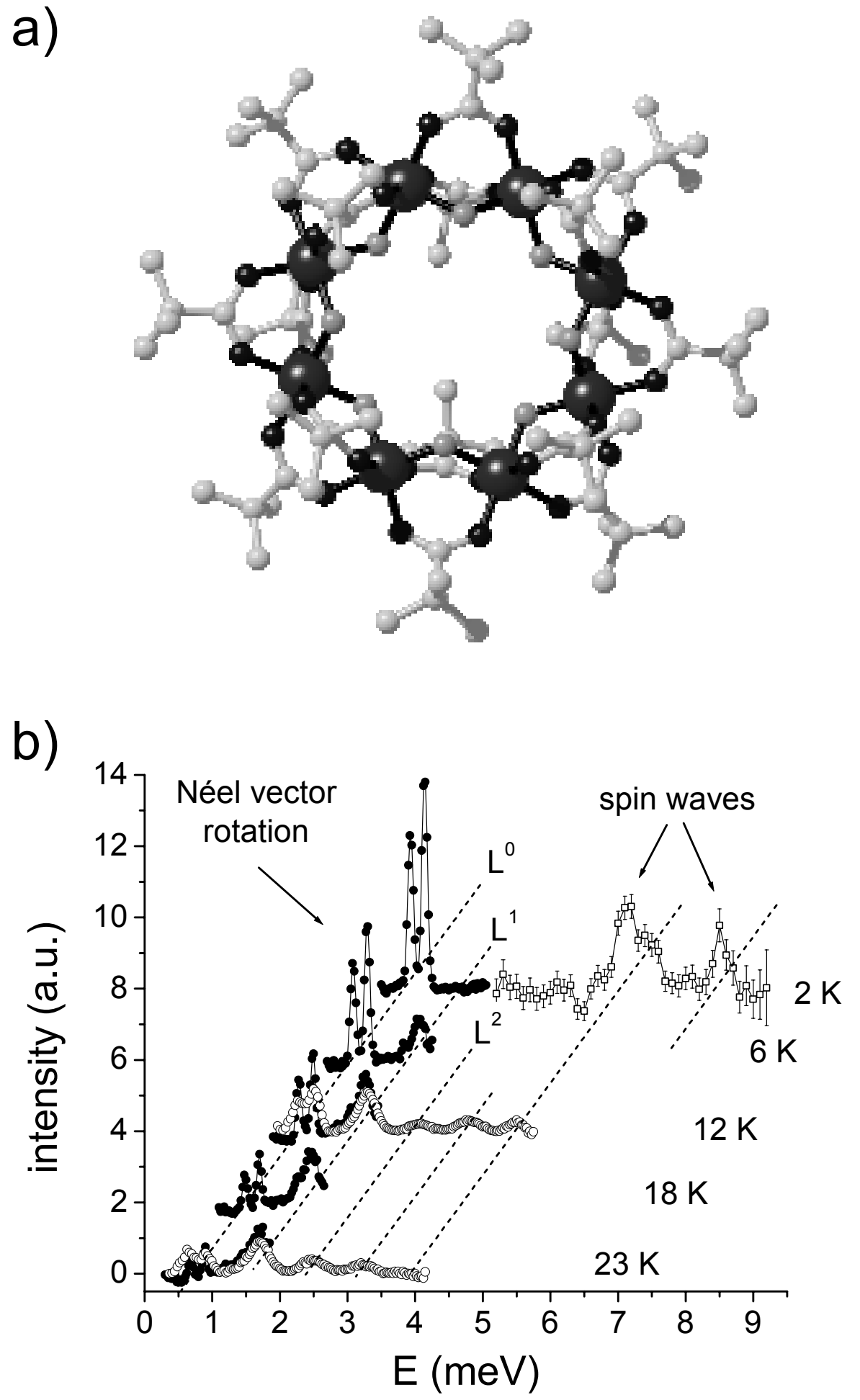

Fig. 10. (a) Structural representation of the Cr8 wheel $[Cr_8F_8\{O_2CC(CH_3)_3\}_{16}]$ (**12**). H atoms are omitted. (b) Inelastic neutron scattering intensity versus energy transfer for Cr8 at different temperatures. Adapted from ref. [48].

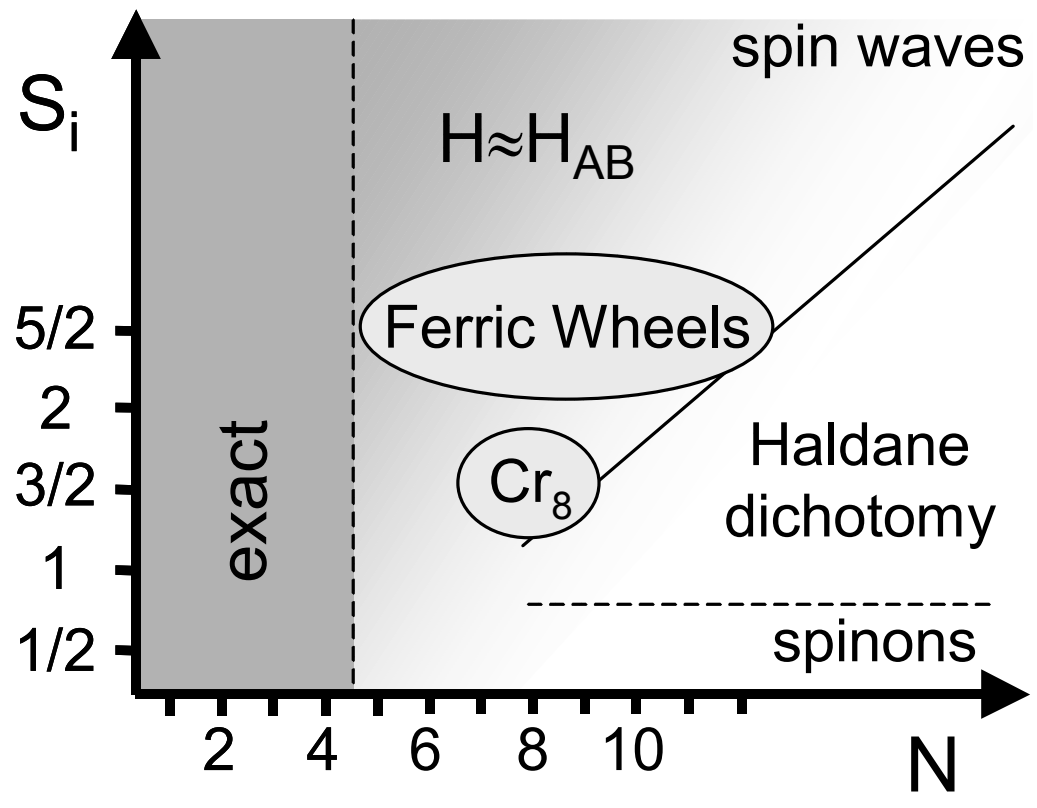

Fig. 11. Graphical overview of the properties of the anti-ferromagnetic Heisenberg ring (AFHR) in the parameter space spanned by the spin length s and the ring size N. The gray shading indicates the area where the spin structure of the AFHR is essentially classical and well described by the spin wave theory and/or the effective spin Hamiltonian (6).

a)

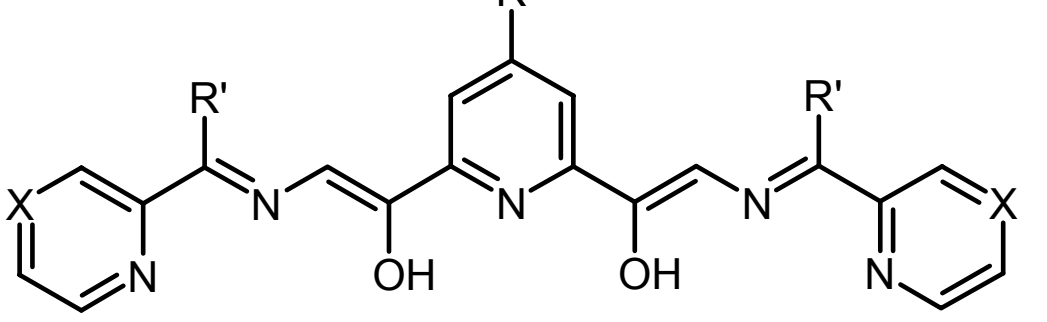

**13a** 2POAP, X = CH, R' = $NH_2$, R = H
**13b** Cl2POAP, X = CH, R' = $NH_2$, R = Cl
**13c** m2POAP, X = CH, R' = $NH_2$, R = OMe
**13d** S2POAP, X = CH, R' = $NH_2$, R = $S^-NH_4^+$
**13e** 2POAPZ, X = N, R' = $NH_2$, R = H
**13f** Cl2POAPZ, X = N, R' = $NH_2$, R = Cl
**13g** 2POMP, X = CH, R' = Me, R = H
**13h** 2POPP, X = CH, R' = Ph, R = H

b)

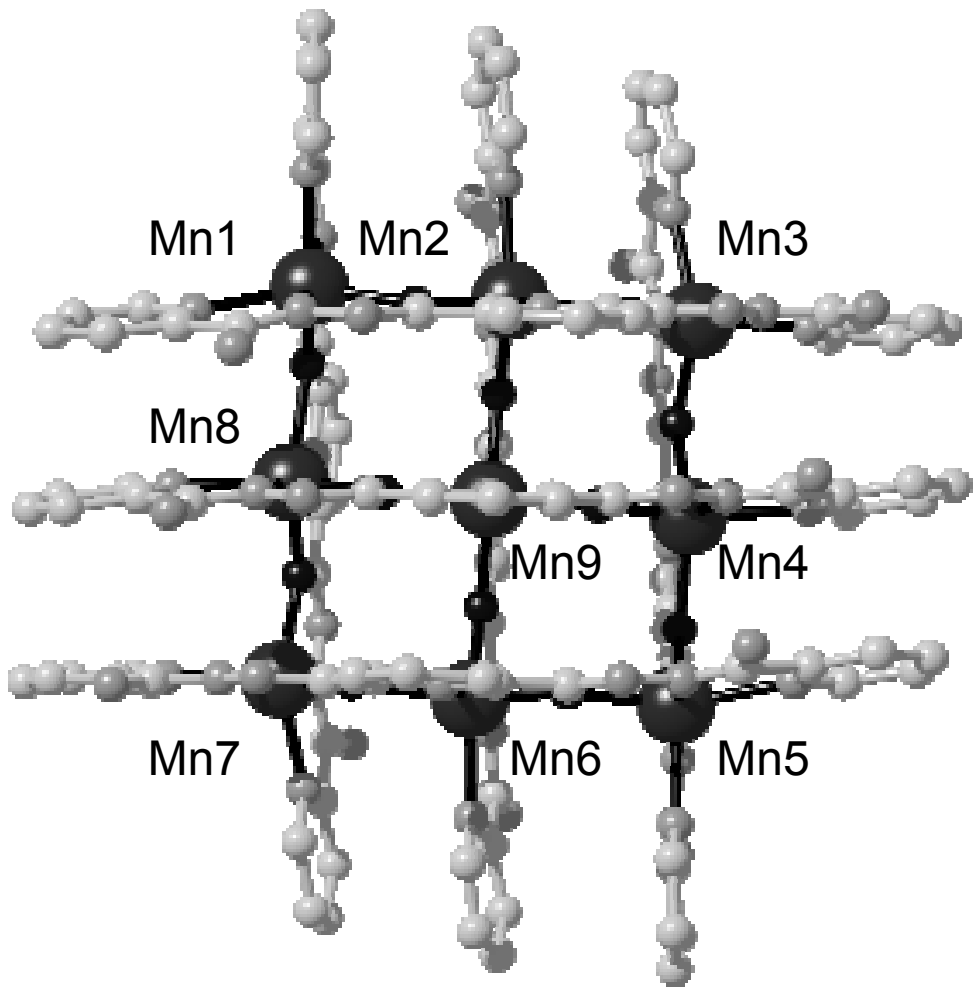

Fig. 12. (a) Sketch of the ligand 2POAP, and (b) structural representation of the Mn(II)-[3 × 3] grid $[Mn_9(\mathbf{13a}\text{-}2H)_6]^{6+}$ (**15**). H atoms are omitted.

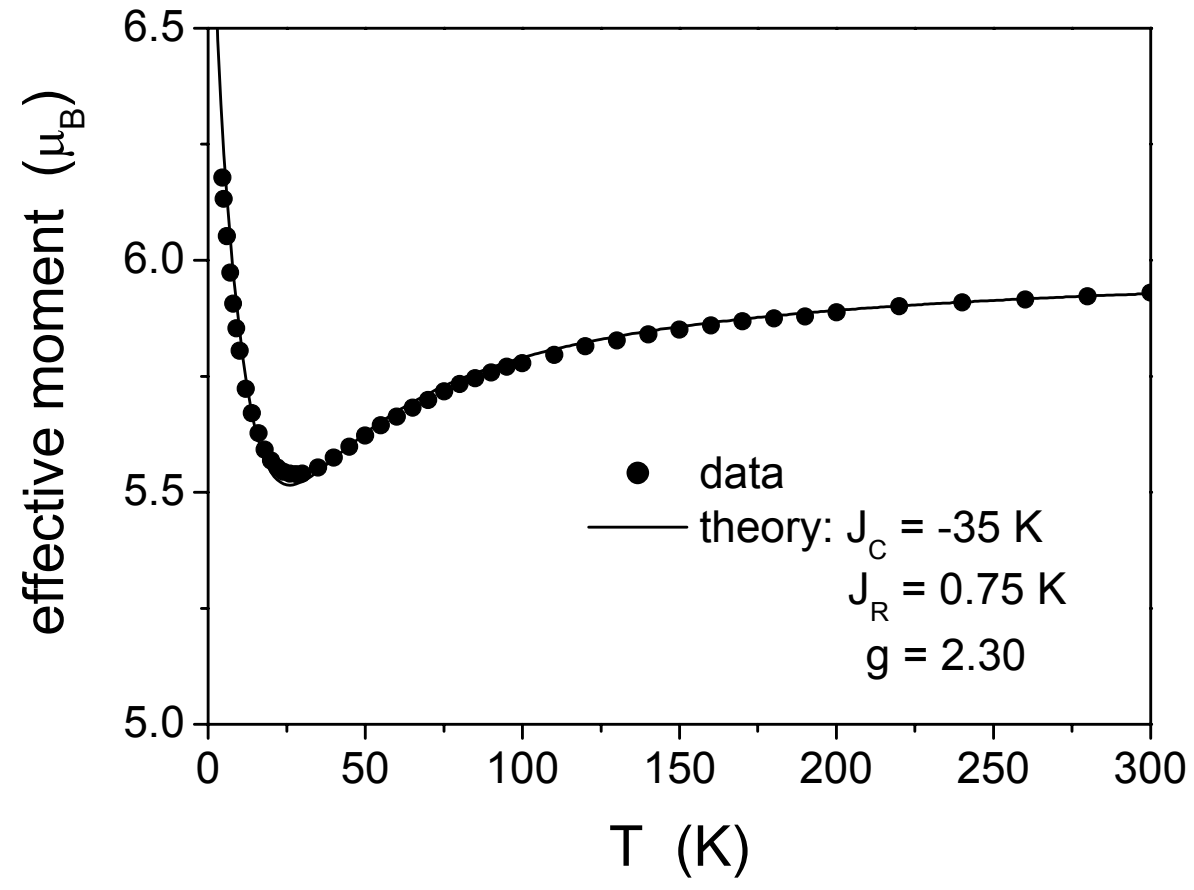

Fig. 13. Temperature dependence of the effective magnetic moment of a powder sample of the Cu(II)-[3 × 3] grid $[Cu_9(\mathbf{13a}\text{-}H)_6](NO_3)_{12}\cdot 9H_2O$ (**14**). Adapted from ref. [58].

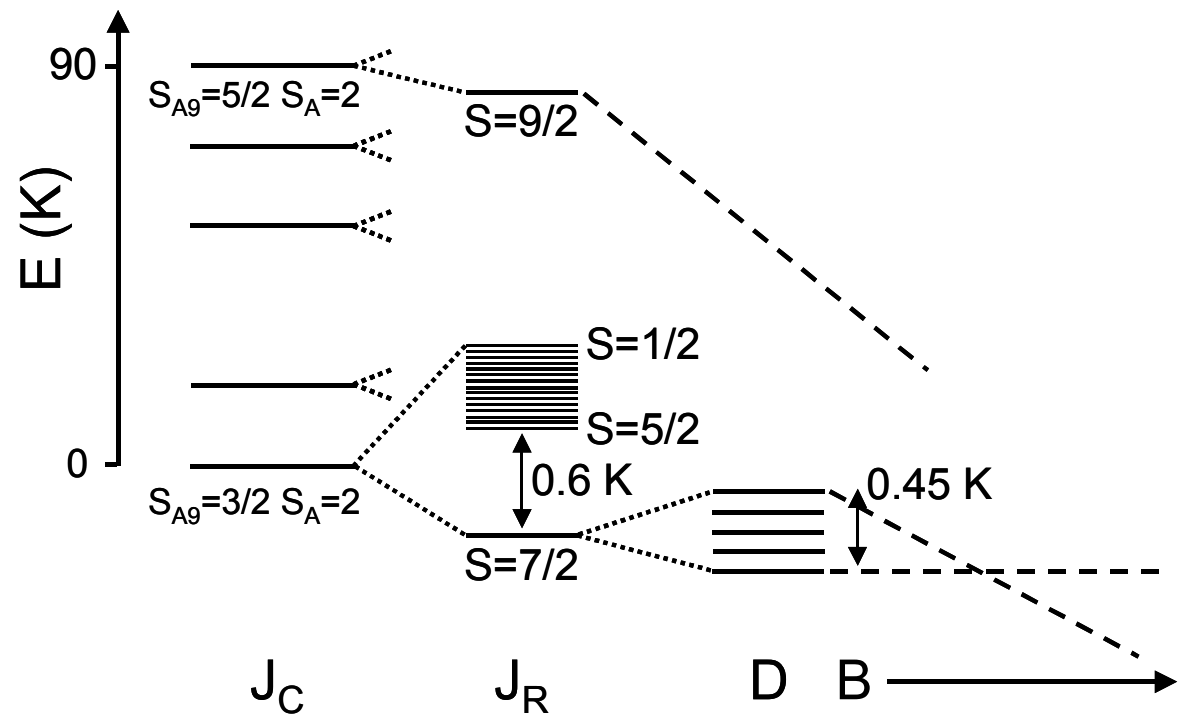

Fig. 14. Schematic drawing of the energy spectrum of the Cu(II)-[3 × 3] grid **14**. The dominating coupling $J_C$ leads to the level pattern of a pentanuclear "star" (given labels correspond to the coupling scheme $\hat{\mathbf{S}}_A = \sum_{i=2,4,6,8} \hat{\mathbf{S}}_i$ , $\hat{\mathbf{S}}_{A9} = \hat{\mathbf{S}}_A + \hat{\mathbf{S}}_9$ ). These spin levels are further split by the weaker couplings $J_R$ resulting in a S = 7/2 ground state, which exhibits a zero-field splitting due to anisotropic exchange and dipole-dipole interactions. In a strong magnetic field, the M = ±7/2 state becomes well separated and is exclusively thermally populated at low temperatures.

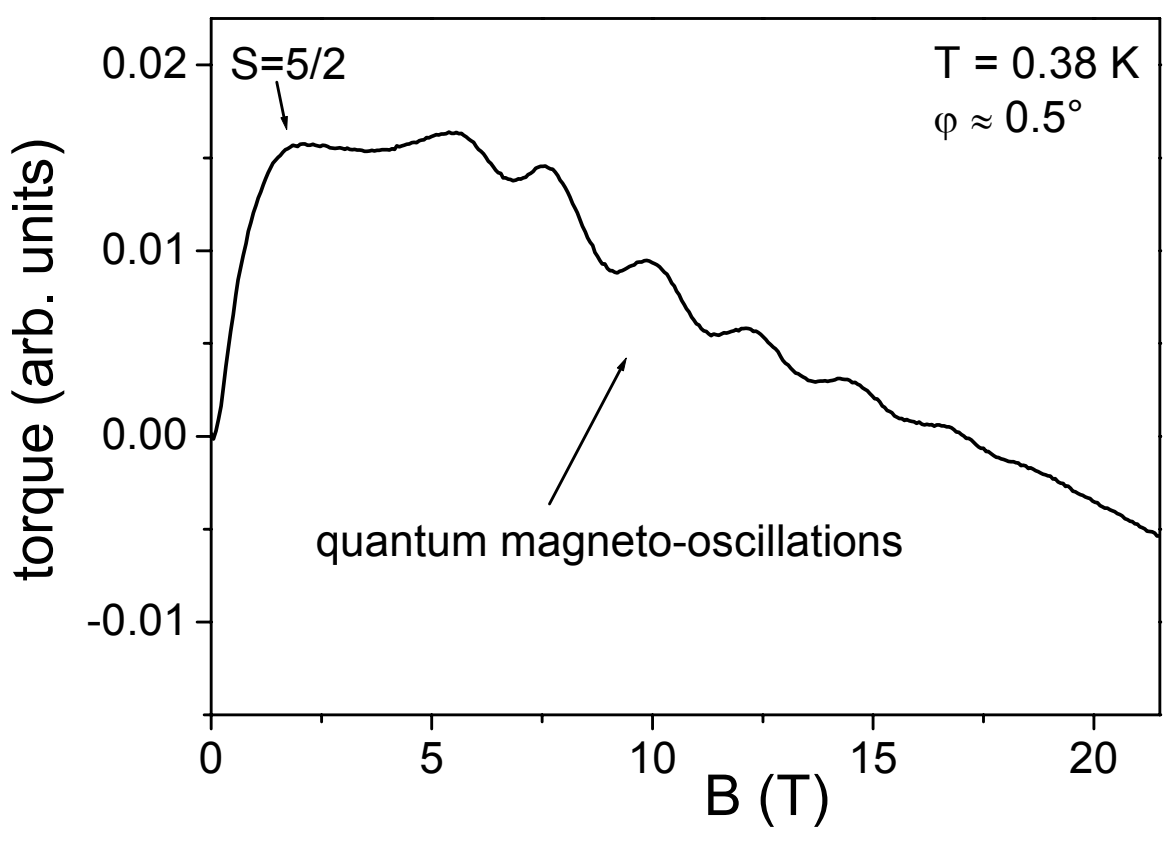

Fig. 15. Torque versus magnetic field of a single crystal of the Mn(II)-[3 × 3] grid $[Mn_9(\mathbf{13a}\text{-}2H)_6](ClO_4)_6\cdot 3.75CH_3CN\cdot 11H_2O$ (**15**) at low temperature. The peak at ≈ 2 T stems from the zero-field splitting of the S = 5/2 ground state. At higher fields, the torque curve exhibits unprecedented quantum magneto-oscillations. Adapted from ref. [60].

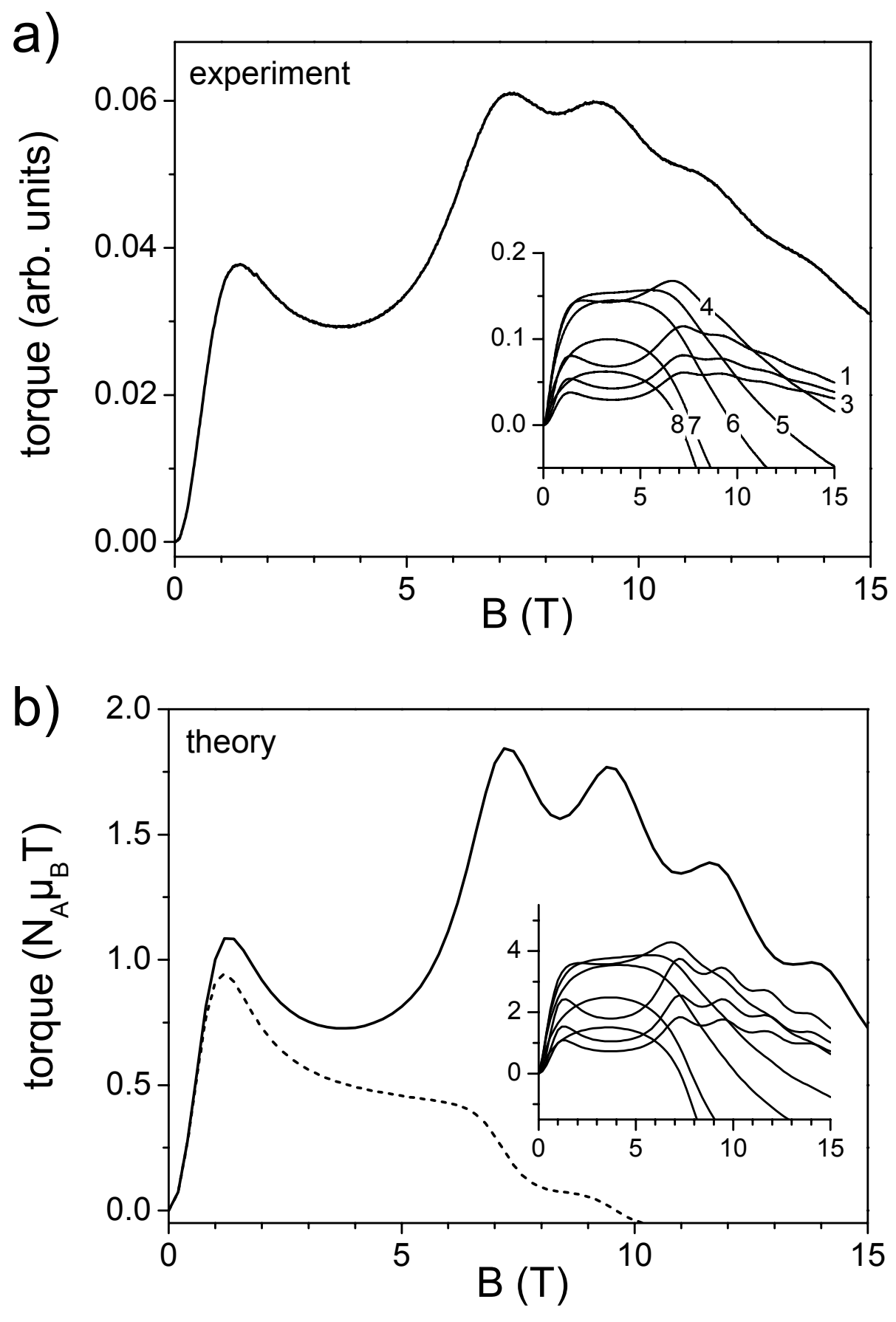

Fig. 16. (a) Experimental and (b) calculated torque curves as functions of the magnetic field of a Mn(II)-[3 × 3] single crystal (**15**) at 0.4 K for various orientations of the magnetic field. The angles between the field and the grid plane were (1) 2.8°, (2) 4.1°, (3) 7.3°, (4) 20.3°, (5) 29.8°, (6) 38.8°, (7) 58.8°, and (8) 70.3°. The main panel shows the curves for an angle of 2.8°. The dashed line in (b) represents the calculated torque curve for 0.4 K and 2.8°, but with effects of level mixing artificially forced to zero. Adapted from ref. [60].

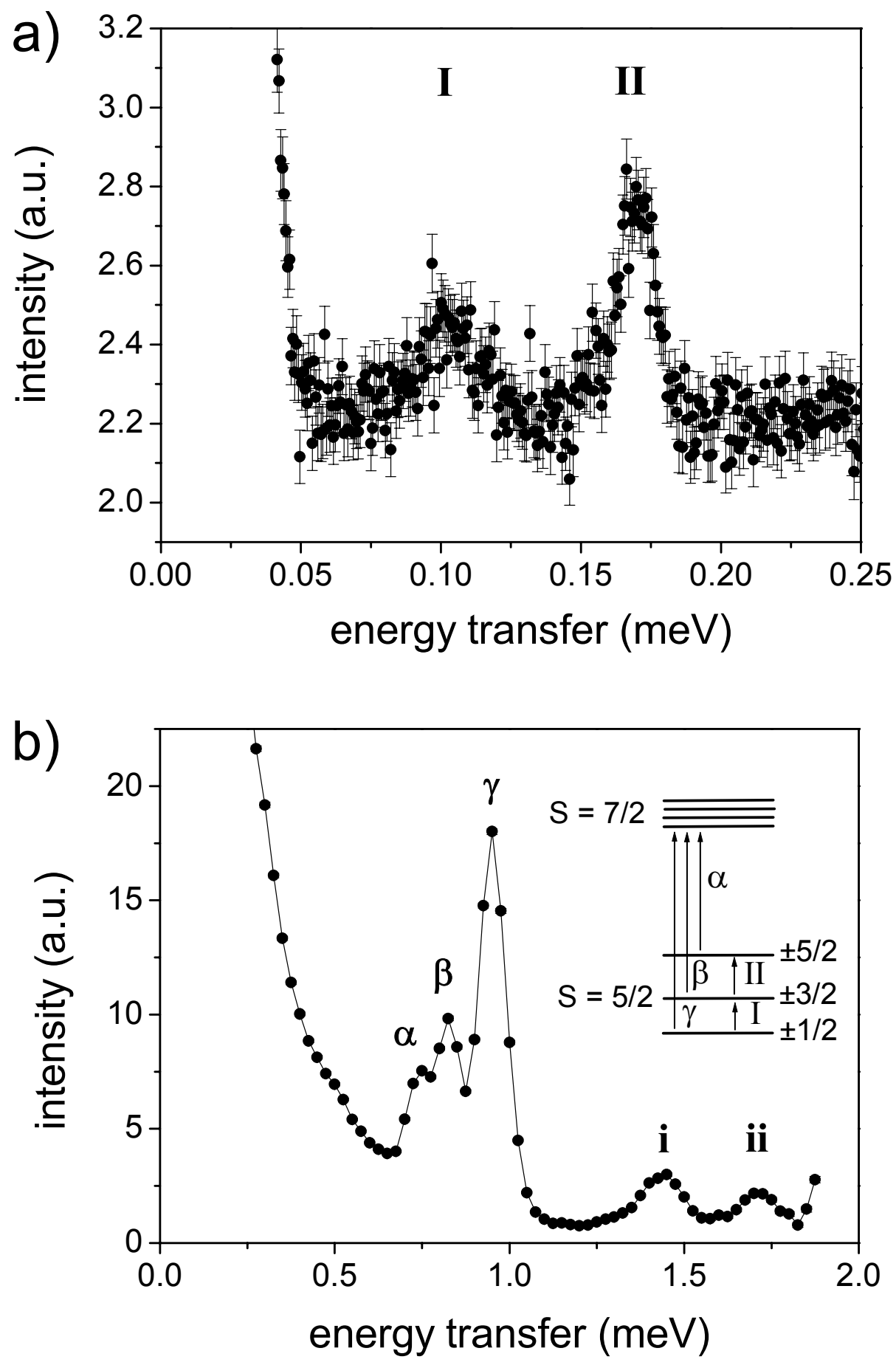

Fig. 17. Inelastic neutron scattering intensity vs. energy transfer for a powder sample of the Mn(II)-[3 × 3] grid **16** at 1.5 K with two different incident neutron energies. Peaks I, II, α, β, and γ correspond to transitions within the S = 5/2 ground-state multiplet and from the S = 5/2 ground state to the first excited S = 7/2 level, respectively, see inset of panel (b). Peaks i and ii correspond to transitions from the S = 5/2 ground state to the E band (compare with Fig. 18). Adapted from ref. [61].

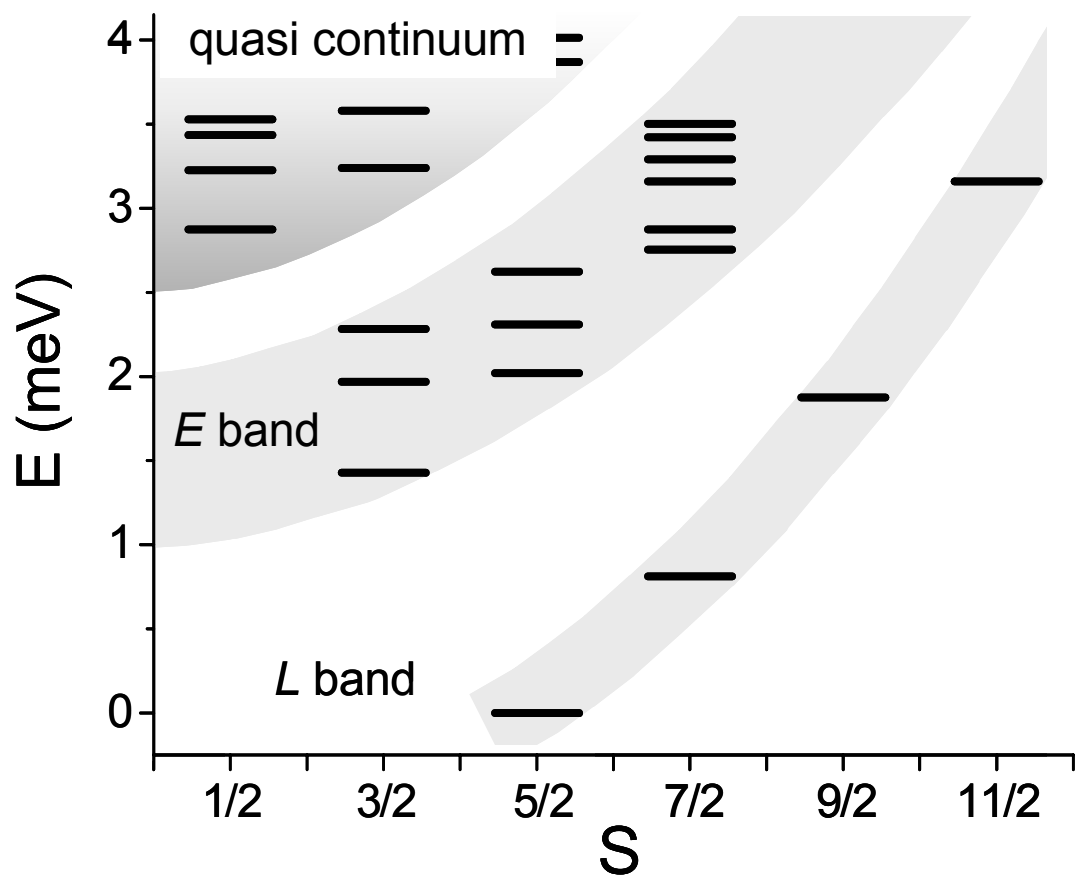

Fig. 18. Low-lying energy spectrum for the Mn(II)-[3 × 3] grid as calculated with the exchange parameters determined from inelastic neutron scattering on **16** (magnetic anisotropy is neglected here).

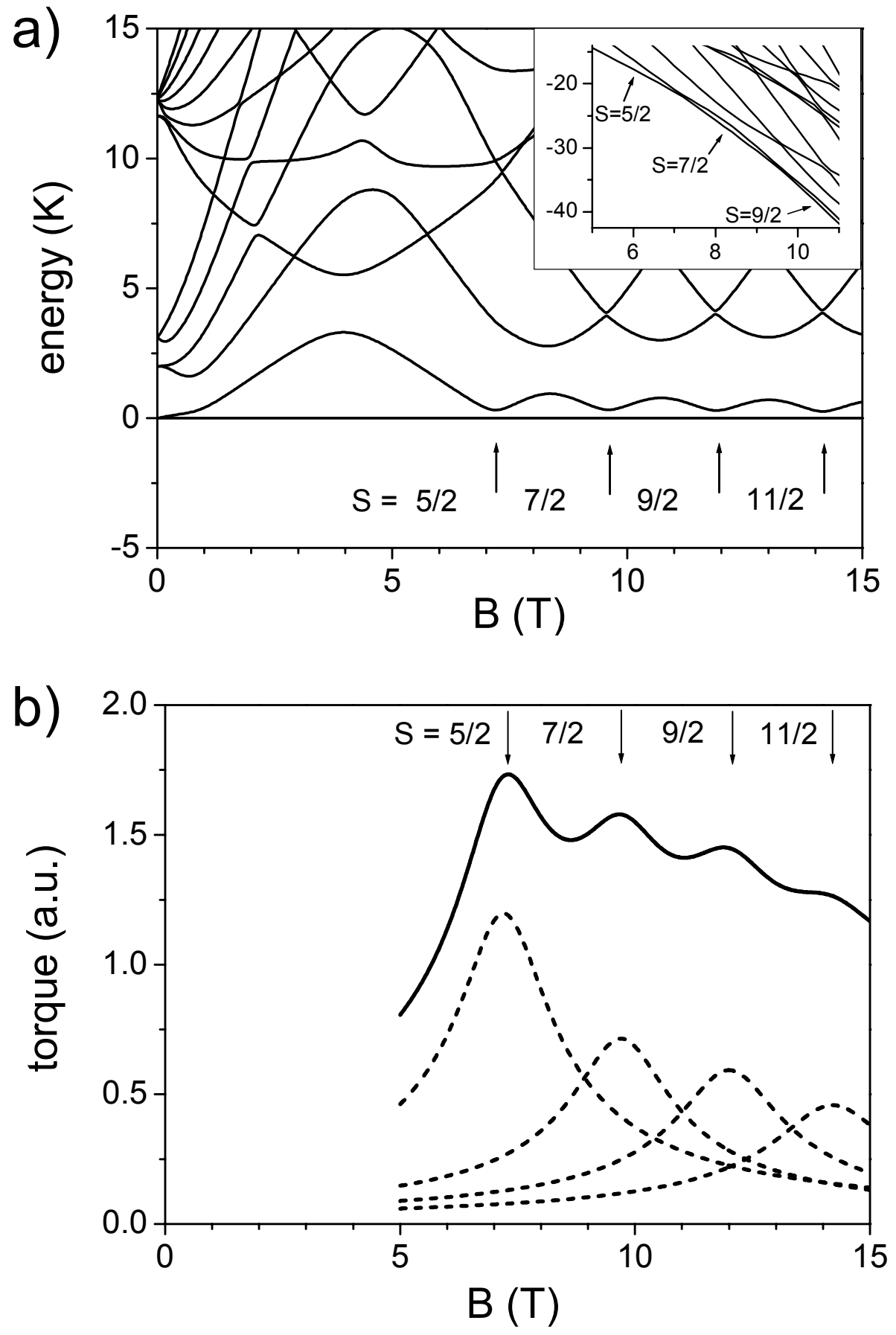

Fig. 19. (a) Calculated energy spectrum vs. magnetic field for Mn(II)-[3 × 3] at an angle of 2.8° (energy of the lowest state was set to zero at each field). Arrows indicate ground-state level crossings, and numbers the total spin S of the respective ground states. The inset details the energy spectrum near the first two level crossings, where the ground state changes as S = 5/2 → 7/2 and S = 7/2 → 9/2 (energies are given here with respect to the ground state energy at zero field). (b) Torque peaks at each level crossing due to the level mixing (dashed lines), which superimpose to produce an oscillatory field dependence (solid line).

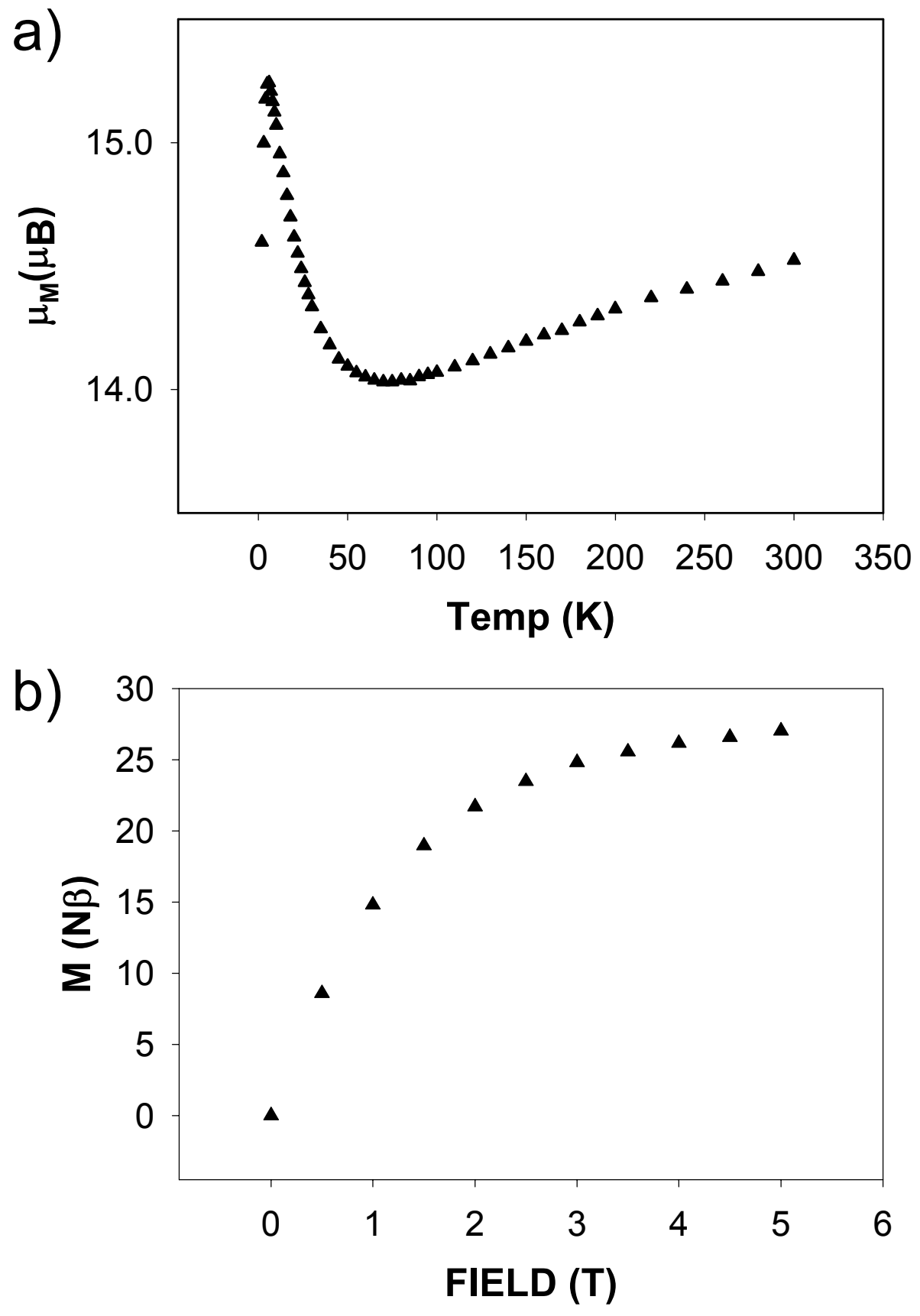

Fig. 20. (a) Temperature dependence of the effective magnetic moment and (b) low temperature magnetization curve of a powder sample of the Fe(III)-[3 × 3] grid $[Fe_9($**13a**$-2H)_6](NO_3)_{15}\cdot 18H_2O$ (**17**). Reprinted with permission from ref. [55]. Copyright (2003) American Chemical Society.

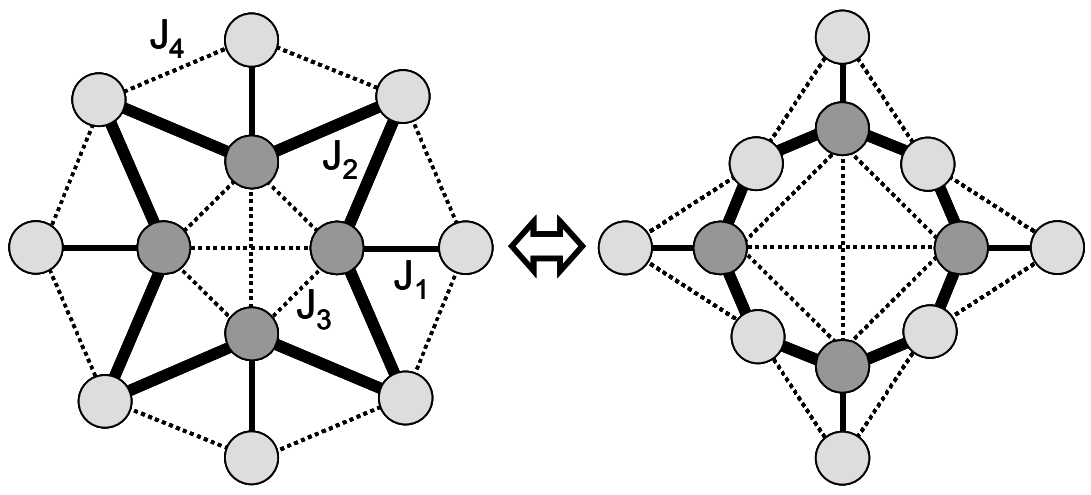

Fig. 21. Exchange coupling paths in the Mn12 molecule as suggested by its structure (on the left). The solid lines indicate the exchange paths $J_1$ and $J_2$, the dashed lines the $J_3$ and $J_4$ interactions. The coupling graph of Mn12 can be equivalently represented as shown on the right. This representation emphasizes the analogy with a ring topology and suggests a classical spin structure in Mn12 due to the dominating $J_1$ and $J_2$ couplings, which is essentially undisturbed by the weak frustration due to $J_3$ and $J_4$.